\documentclass{lmcs} 
\pdfoutput=1

\usepackage{lastpage}
\lmcsdoi{15}{1}{19}
\lmcsheading{}{\pageref{LastPage}}{}{}%
{Oct.~31,~2017}{Mar.~05,~2019}{}


\usepackage{comment}
\usepackage{amsmath}
\usepackage{amsfonts}
\usepackage{amssymb}
\usepackage{xspace}
\usepackage{thmtools}
\usepackage{thm-restate}
\usepackage[numbers,square,semicolon,sectionbib]{natbib}
\usepackage{hyperref}

\usepackage{graphicx}
\usepackage{color}
\usepackage{etoolbox}

\newcommand{\executeiffilenewer}[3]{%
\ifnum\pdfstrcmp{\pdffilemoddate{#1}}%
{\pdffilemoddate{#2}}>0%
{\immediate\write18{#3}}\fi%
} 
\def\cqedsymbol{\ifmmode$\lrcorner$\else{\unskip\nobreak\hfil
\penalty50\hskip1em\null\nobreak\hfil$\lrcorner$
\parfillskip=0pt\finalhyphendemerits=0\endgraf}\fi}

\newcommand{\eff}{\textsc{eff}}
\newcommand{\len}{\textsc{len}}
\newcommand{\pref}{\textsc{pref}}
\newcommand{\midd}{\textsc{midd}}
\newcommand{\cyc}{\textsc{cyc}}
\newcommand{\premidd}{\textsc{pre}}
\newcommand{\postmidd}{\textsc{post}}
\newcommand{\preconn}{\textsc{pre-conn}}
\newcommand{\postconn}{\textsc{post-conn}}
\newcommand{\suff}{\textsc{suff}}

\newcommand{\up}{\textsc{up}}
\newcommand{\down}{\textsc{down}}
\newcommand{\rcap}{\textsc{cap}}

\newcommand{\ctproj}{\textsc{proj}}
\newcommand{\cvalue}{\textsc{cnt}}
\newcommand{\lcm}{\textup{lcm}}
\newcommand{\fasten}{\textsc{fasten}}
\newcommand{\source}{\textsc{src}}
\newcommand{\target}{\textsc{targ}}
\newcommand{\state}{\textsc{st}}
\newcommand{\low}{\textsc{low}}

\newcommand{\sign}{\textsc{sign}}

\newcommand{\SCCs}{\mathfrak{S}}
\newcommand{\Reg}{\mathcal{N}}
\newcommand{\Low}{\mathcal{L}}

\newcommand{\N}{\mathbb{N}}
\newcommand{\Z}{\mathbb{Z}}
\newcommand{\trans}[1]{\stackrel{#1}{\longrightarrow}}

\newcommand{\MyOmega}{\mathrm{\Omega}}
\newcommand{\MySigma}{\mathrm{\Sigma}}
\newcommand{\sset}{\subseteq}
\newcommand{\eps}{\varepsilon}

\newcommand{\newextmathcommand}[2]{%
    \newcommand{#1}{\ensuremath{#2}\xspace}
}

\newextmathcommand{\OCA}{\mathcal A}
\newextmathcommand{\OCS}{\mathcal O}
\newcommand{\const}{\mathrm{const}}
\newcommand{\polylog}{\mathrm{polylog}}

\excludecomment{Append}
\excludecomment{NewAppend}

\begin{document}

\title[Shortest paths in one-counter systems]{Shortest paths in one-counter systems}
\titlecomment{{\lsuper*}This is the full version of the paper~\cite{ChistikovCHPW16} that appeared
in the proceedings of FoSSaCS'15.
The main part of these results was obtained during Autob\'oz'15,
the annual research camp of Warsaw automata group. During the work on these
results, D.~Chistikov was a postdoctoral researcher at the
Max Planck Institute for Software Systems (MPI-SWS) in Kaiserslautern and Saarbr\"ucken, Germany, and
Mi.~Pilipczuk held a post-doc position at Warsaw Centre of Mathematics
and Computer Science and was supported by the Foundation for Polish Science via
the START stipend programme. P.~Hofman was supported by 
Labex Digicosme, Univ. Paris-Saclay, project
VERICONISS. W.~Czerwi\'{n}ski acknowledges a partial support by the Polish National
Science Centre grant 2013/09/B/ST6/01575. M.~Wehar was a PhD student for the Department of Computer Science and Engineering at University at Buffalo.}

\author[D.~Chistikov]{Dmitry Chistikov\rsuper{a}}	
\address{\lsuper{a}Centre for Discrete Mathematics and its Applications (DIMAP) \& Department of Computer Science, University of Warwick, UK}	
\email{d.chistikov@warwick.ac.uk}  

\author[W.~Czerwi\'nski]{Wojciech Czerwi\'nski\rsuper{b}}	
\address{\lsuper{b}Institute of Informatics, University of Warsaw, Poland}	
\email{\{wczerwin,ph209519,michal.pilipczuk\}@mimuw.edu.pl}  

\author[P.~Hofman]{Piotr Hofman\rsuper{b}}	
\address{\lsuper{c}Computer \& Information Sciences, Temple University, USA}	
\email{michael.wehar@temple.edu}  

\author[M.~Pilipczuk]{Micha{\l} Pilipczuk\rsuper{b}}	
\author[M.~Wehar]{Michael Wehar\rsuper{c}}	

\begin{abstract}
\noindent We show that any one-counter automaton with $n$ states, if its language is
non-empty, accepts some word of length at most $O(n^2)$.
This closes the gap between the previously known upper bound of $O(n^3)$ and
lower bound of $\MyOmega(n^2)$.
More generally, we prove a tight upper bound on the length of shortest paths
between arbitrary configurations in one-counter transition systems (weaker bounds have previously appeared in the literature).
\end{abstract}

\maketitle

\section{Introduction}

\newcommand{\MyTheta}{\mathrm{\Theta}}
\newcommand{\df}[1]{\emph{#1}}
\newcommand{\poly}{\mathrm{poly}}

Extremal combinatorial questions are ubiquitous in today's
theory of computing:
How many steps does an algorithm take in the worst case
when traversing a data structure?
How large is the most compact automaton for a formal language?
While some specific questions of this form are best seen as standalone puzzles,
only interesting for their own sake,
others can be used as basic building blocks for
more involved arguments.

We look into the following extremal problem:
Given a one-counter automaton \OCA with $n$ states, how long can
the shortest word accepted by \OCA be?
It is folklore that, unless the language of \OCA is empty,
\OCA accepts some word of length at most polynomial in~$n$.
This fact and a number of related results of similar form have appeared
as auxiliary lemmas in the literature on formal languages,
analysis of infinite-state systems, and applications of formal methods~%
\citetext{%
    \citealp[Lemma~6]{LafourcadeLT05};
    \citealp[Section~8.1]{LafourcadeLT05tr};
    \citealp[Lemma~5]{EtessamiWY10};
    \citealp[Lemma~11]{AlurC11};
    \citealp[Lemmas~28 and~29]{HofmanMT13};
    \citealp[Section~5]{DemriG09}%
}.
A closer inspection reveals
that the arguments behind these results deliver (or can deliver) an upper bound of $O(n^3)$,
while the best known lower bound comes from examples of one-counter automata
with shortest accepted words of length $\MyTheta(n^2)$.
In other words, the true value is at least quadratic and at most cubic.

The main result of the present paper is that
we close this gap by showing a quadratic upper bound, $O(n^2)$.
This upper bound was previously conjectured by Wojtczak~\citep[Conjecture~3.3.4]{W-thesis}.
We also extend this result to a more general reachability setting:
in any one-counter (transition) system with $n$ control states,
whenever there is a path from a configuration $\alpha$
to a configuration $\beta$---recall that configurations are
pairs of the form $(q, c)$ where $q \in Q$ is the control state,
$|Q| = n$, and $c$ is a counter value, a nonnegative integer---%
there is also a path from $\alpha$ to $\beta$ that has length
at most $O(n^2 + n \cdot \max (c_\alpha, c_\beta))$ where $c_\alpha$ and $c_\beta$
are the counter values of $\alpha$ and $\beta$.
We discuss our contribution in more detail in the following
Section~\ref{s:overview}.

During the review process of (the journal version of) this paper,
we became aware of the 1986 paper by Del\'eage and Pierre~\cite{DP86}
that shows an upper bound of $2 n^2 + 4 n$ on the rational index
of the language of balanced parentheses (the restricted Dyck language $D_1'^*$).
This is an upper bound on the length of shortest words accepted
by one-counter automata that have no zero tests and accept by
final state and zero counter value.
Upper bounds in our work apply in a more general setting,
where automata may have zero tests (and, less importantly, accept
by final state only). There appears to be no suitable reduction from
this problem to the special case without zero tests; the natural link
between the problems only yields an upper bound of $n \cdot (2 n^2 + 4 n) = O(n^3)$.

\medskip
\subsubsection*{Related work and motivation.}
Reachability is a fundamental problem in theoretical computer science and in its applications in verification,
notably via analysis of infinite-state systems~\citep{BouajjaniEM97,Thomas09,AtigBKS14,LerouxS15}.
Among such systems, counter-based models of computation are a standard abstraction
that has attracted a lot of attention~\citep{BarrettDD13,FarzanKP14,BouajjaniHM03};
machines with a single counter are, of course, the most basic.
While our main motivation has been purely theoretical,
we note that bounds on the length of shortest paths
in one-counter systems have appeared as building blocks
in the literature on rather diverse topics.

More specifically,
a polynomial upper bound 
is used by Etessami, Wojtczak, and Yannakakis~\citep{EtessamiWY10}
and Stewart, Etessami, and Yannakakis~\citep{StewartEY15} in an analysis of
\emph{probabilistic} one-counter systems (which are equivalent to so-called
discrete-time quasi-birth-death processes, QBDs).
Etessami et~al.~\cite{EtessamiWY10} prove that in the $(q, 1) \leadsto (q', 0)$-reachability setting
the counter does not need to grow higher than $n^2$ and provide examples
showing that this bound is tight.
However, they only deduce upper bounds of $n^3$ and $n^4$ on the length of
shortest paths without and with zero tests, respectively.
A simple corollary shows that if a state $q$ can eventually reach a state $q'$
with a non-zero probability, then this probability is lower-bounded by $p^{\,\poly(n)}$
where $p$ is the smallest among positive probabilities associated with transitions.
This becomes a step in the proof that a (decomposed) Newton's method
approximates \emph{termination probabilities} of the system
in time polynomial in its size, $n$:
both for the unit-cost rational arithmetic RAM~\citep{EtessamiWY10}
and for the Turing model of computation~\citep{StewartEY15}.
The results of the present paper
prove a conjecture stated by Wojtczak~\citep[Conjecture~3.3.4]{W-thesis} and
reduce the (theoretical) worst-case
upper bounds on the number of steps roughly by a factor of $n$.

In a subsequent work, Hofman et~al.~\citep{HofmanMT13} reuse the auxiliary lemmas
on the length of shortest paths from~\citep{EtessamiWY10}
and show that (strong and weak) trace inclusion for a one-counter system and
a finite-state process is decidable in PSPACE (and is, in fact, PSPACE-complete).

One may note that a stronger upper bound of $O(n^3)$ on the length of shortest paths
can be derived from the above bound on the largest needed counter value
even in the presence of zero tests.
This value, $O(n^3)$, seems to be a recurring theme in the literature on one-counter systems;
it already appears in the pumping lemma for one-counter languages
due to Latteux~\citep{Latteux83} as the \df{pumping constant}:
a number $N$ such that any accepted word longer than $N$ can be pumped.
In fact, the formulation in~\citep{Latteux83} does not permit \emph{removals}
of factors from an accepted word, but even such a version would only yield
the same upper bound of $O(n^3)$ on the length of shortest paths.
While the arguments of the present paper do not lead to an improvement in
the pumping constant for one-counter languages (see Section~\ref{s:pumping-constant}),
we nevertheless show that in the reachability setting
the optimal value (the length of the shortest path) is actually $O(n^2)$.

A cubic upper bound on the largest needed counter value
(for the reachability setting) in one-counter systems
without zero tests, also known as one-counter nets, appears in
the work of Lafourcade et~al.~\citep{LafourcadeLT05,LafourcadeLT05tr}.
This result is applied in the context of the Dolev-Yao intruder model,
where the question of whether a passive eavesdropper (an intruder) can obtain
a piece of information is reduced to the decision problem
for a deduction system.
For several such systems, Lafourcade et~al. show that,
under certain assumptions,
the problem is decidable in polynomial time.
They construct a one-counter system where
states represent terms from a finite set and the counter value
corresponds to the number of applications of a free unary function symbol
to a term. After this, the upper bound on counter values along shortest paths
is extended to an upper bound on the size of terms that can be used
in a minimal deductive proof;
needless to say, an improvement in the upper bound extends in a natural way.

Finally, we would like to mention the work of Alur and \v{C}ern\'y~\citep{AlurC11},
who use a related model of one-counter systems with counter values in~$\mathbb Z$
and without zero tests.
They reduce the equivalence problem for so-called streaming data-string
transducers to $(q, 0) \leadsto (q', 0)$-reachability in such counter systems:
the transducers produce output at the end of the computation, and
the counter is used to track the accumulated distance between
a distinguished pair of symbols in the output.
Since these transducers are designed to model list-manipulating programs
(in two syntactically restricted models),
decision procedures for equivalence of such programs
can rely on the upper bounds for shortest paths to efficiently prune
the search space.
In~\cite{AlurC11}, the upper bound on the path length is the familiar $O(n^3)$;
this gives an upper bound on the length of smallest counterexamples to equivalence.
Our upper bound of $O(n^2)$ extends to this model of counter systems too.
(An equivalent question appears in the work of
Ang et~al.~\cite[Propositions~6 and~7]{AngPRS09},
also with a quadratic lower bound and cubic upper bound.)
The reduction to reachability in one-counter systems
was recently implemented by Thakkar et~al.~\cite{ThakkarKA13}
on top of ARMC, an abstraction-refinement model checker~\cite{ARMC},
for the purpose of verifying retransmission protocols over noisy channels.


\section{Summary}
\label{s:overview}

\subsubsection*{One-counter systems.}
In this paper we work in the framework of one-counter systems,
which are an abstract version of one-counter automata.
More precisely, they are one-counter automata without input alphabet
(see below).

Formally, a \emph{one-counter system (OCS)} \OCS consists of a finite set of states $Q$,
a set of non-zero transitions $T_{>0} \subseteq Q \times \{-1,0,1\} \times Q$,
and a set of zero tests $T_{=0} \subseteq Q \times \{0, 1\} \times Q$.
A \emph{configuration} of the OCS~\OCS is a pair in $Q \times \N$.
We define a binary relation $\trans{}$ on the set $Q \times \N$
as follows: $(p, c) \trans{} (q, c+d)$ whenever
(i) $c \ge 1$ and $(p, d, q) \in T_{>0}$ or
(ii) $c = 0$ and $(p, d, q) \in T_{=0}$.
The reflexive transitive closure of $\trans{}$ is denoted by $\trans{}^*$;
we say that a configuration $\beta$
is \emph{reachable} from $\alpha$ if $\alpha \trans{}^* \beta$.
This reachability is witnessed by a \emph{path} in OCS~\OCS,
which is simply a path in the infinite directed graph with vertices $Q \times \N$ and
edge relation $\trans{}$; vertices and edges along the path can be repeated.
The \emph{length} of the path is the number of (not necessarily distinct) edges
that occur on it.

\medskip
\subsubsection*{Our contribution.}
We first formulate our results in terms of one-counter systems.
Our first result is on paths between configurations with zero
counter~values.

\begin{thm}\label{thm:main-thm}
Let \OCS be a one-counter system with $n$ states.
Suppose a configuration $\beta=(p_\beta,0)$ is reachable
from a configuration $\alpha=(p_\alpha,0)$ in~\OCS.
Then \OCS~has a path from $\alpha$ to $\beta$
of length at most $14 n^2$.
\end{thm}

\noindent
Using Theorem~\ref{thm:main-thm} as a black-box, we generalize it to
the case where the source and target configurations have arbitrary counter values.

\begin{restatable}[]{thm}{restategeneralization}\label{thm:generalization}
Let \OCS be a one-counter system with $n$ states.
Suppose a configuration $\beta=(p_\beta,c_\beta)$ is reachable
from a configuration $\alpha=(p_\alpha,c_\alpha)$ in \OCS.
Then \OCS~has a path from $\alpha$ to $\beta$
of length at most $14 n^2 + n \cdot \max(c_\alpha,c_\beta)$.
\end{restatable}

\noindent
The proof of Theorem~\ref{thm:main-thm} is the main technical contribution of this work.
We prove Theorem~\ref{thm:main-thm} in Section~\ref{s:proof}
and Theorem~\ref{thm:generalization}, as well as an extension
to OCS with negative counter values, in Section~\ref{s:generalizations}.

\medskip
\subsubsection*{One-counter automata.}
We now restate our contribution in terms of one-counter automata
(which are the original motivation for this work).

Take any finite set $\MySigma$. The set of all finite words over $\MySigma$
is denoted by $\MySigma^*$, and the empty word by $\eps$.
A (\emph{nondeterministic}) \emph{one-counter automaton (OCA)}~\OCA
over the input alphabet $\MySigma$ is a one-counter system
where every transition $t \in T_{>0} \cup T_{=0}$ is associated
with a label, $\lambda(t) \in \MySigma \cup \{\eps\}$, and
where some subsets $I \sset Q$ and $F \sset Q$ are distinguished as
sets of initial and final states respectively.
The labeling function $\lambda$
is extended from transitions to edges $\trans{}$ and to paths
in a natural way;
the automaton \emph{accepts} all words that are labels of paths from
$I \times \{0\}$ to $F \times \N$.
The \emph{language} of a one-counter automaton~\OCA
is the set $L \sset \MySigma^*$ of all words accepted by~\OCA.

\begin{cor}
\label{cor:automata}
Let \OCA be a nondeterministic one-counter automaton with $n$ states.
If the language of~\OCA is non-empty, then \OCA
accepts some word of length at most~$14 n^2$.
\end{cor}

\begin{proof}
Take \OCA with a non-empty language and add
self-loops with decrements to all final states:
$(p, -1, p) \in T_{>0}$ for $p \in F$.
Since the language of~\OCA is non-empty,
some final configuration (in $F \times \N$) is reachable
from some initial configuration (from $I \times \{0\}$);
this implies that in the modified automaton, denoted by~$\OCA'$,
a configuration $\beta = (p_\beta, 0)$, $p_\beta \in F$, is reachable
from a configuration $\alpha = (p_\alpha, 0)$, $p_\alpha \in I$.
Consider the shortest path in~$\OCA'$ between $\alpha$ and $\beta$:
by Theorem~\ref{thm:main-thm}, its length is at most $14 n^2$.
Take the shortest prefix of this path that contains a state from $F$;
this path is a path in~\OCA.
Since the label of the path cannot be longer than
the path itself, the result follows.\end{proof}

As a concrete example,
from Corollary~\ref{cor:automata} it follows that
any nondeterministic one-counter automaton that accepts the singleton
unary language $\{ a^n \}$ ---a basic version of counting to~$n$--- must
have at least $\MyOmega(\sqrt n)$ states.
This lower bound is tight and shows that nondeterminism does not help
to ``count to~$n$'', because
\emph{deterministic} one-counter automata can also do this
using $\MyTheta(\sqrt n)$ states~\cite{Chistikov14}.

\medskip
\subsubsection*{Lower bounds.}
As we already said, the lower bound on the length of the shortest
path is $\MyOmega(n^2)$.
We present constructions of OCS that match the upper bounds of
Theorems~\ref{thm:main-thm} and~\ref{thm:generalization}.
Note that Examples~\ref{ex:low-counter} and~\ref{ex:high-counter}
seem to use different phenomena.

\begin{exaC}[{\cite{EtessamiWY10,Chistikov14}}]\label{ex:low-counter}
Consider an OCS~$\OCS_1$ with $2 n$ states: $p_1, \ldots, p_n$ and $q_1, \ldots, q_n$.
Let~$\OCS_1$ have, for $1 \leq i < n$, transitions $(p_i, +1 ,p_{i+1})$ and
$(q_i, 0, q_{i+1})$, as well as $(q_n, -1,q_1)$ and $(p_n,0,q_1)$.
All the transitions are non-zero, except for transition $(p_1, +1, p_2)$,
which is a zero test.
This OCS is deterministic: every configuration
has at most one outgoing transition.
The only path from $(p_1,0)$ to $(q_1,0)$ has length $n^2$.
\end{exaC}

\begin{exaC}[{\cite{DP86,EtessamiWY10}}]\label{ex:high-counter}
Let $k$ and $m$ be coprime
and let the OCS~$\OCS'_2$ have states
$p_0, \ldots, p_{k-1}$, $q_0, \ldots, q_{m-1}$, and $s_1, s_2$.
Let $\OCS'_2$ have,
for all $0 \leq i < k$ and $0 \leq j < m$,
non-zero transitions $(p_i, +1, p_{i+1 \bmod k})$ and
$(q_j, -1, q_{j+1 \bmod m})$, a non-zero $(p_0,-1,q_1)$, and
zero tests $(s_1,+1,p_1)$, $(q_0,0,s_2)$.
Now paths from $(s_1,0)$ to $(s_2,0)$ correspond to solutions
of $x \cdot k - y \cdot m  = 0$; the shortest path takes
the first cycle $x = m$ times and the second cycle $y = k$ times.
Exiting the second cycle uses an additional transition, making the length $2 k m + 1$.
Setting $k = n$ and $m = n - 1$ gives an OCS~$\OCS_2$
with $2 n + 1$ states where not only does the shortest path
have quadratic length, but all such paths also need to use
quadratic counter values.
\end{exaC}

\begin{exa}
This example justifies the need for the term $n \cdot \max(c_\alpha,c_\beta)$
in Theorem~\ref{thm:generalization}.
Modify $\OCS_1$ from Example~\ref{ex:low-counter} as follows.
Add states $a_1, \ldots, a_n$, $b_1, \ldots, b_n$
and the following non-zero transitions: $(a_n,-1,a_1)$, $(b_n,+1, b_1)$,
and, for all $0 \le i < n$, $(a_i,0,a_{i+1})$ and $(b_i,0,b_{i+1})$. For each of these non-zero transitions, apart from $(a_n,-1,a_1)$, introduce also the same transition as a zero test. Finally, add two more zero tests: $(a_n,0, p_1)$ and $(q_1,0,b_1)$. Thus, the obtained OCS~$\OCS_3$ has $4n$ states.
Observe that every path in~$\OCS_3$ from $(a_1, c_\alpha)$ to $(b_n,c_\beta)$
has to go through $(a_n, 0)$ and $(b_1,0)$
and thus has length at least $n^2 + n (c_\alpha+c_\beta+2)$.
\end{exa}


\section{Challenges and techniques}
\label{s:techniques}

We now discuss shortly the intuition behind our approach to proving
Theorem~\ref{thm:main-thm}, and where the main challenges lie. 

The first, obvious observation is as follows:
if some configuration appears more than once
on a path, then the segment between any two appearances of this configuration
can safely be removed. If we apply this modification exhaustively, then on each
``level'' --- a set of configurations with the same counter value --- we cannot
see more than $n$ configurations. If the maximum counter value observed on some
path were bounded by $O(n)$, then we would immediately obtain a quadratic
upper bound on its length. Unfortunately, this is not the case: as
Example~\ref{ex:high-counter} shows, the counter values in the shortest
accepting path can be as large as quadratic. Hence, applying this observation in
a straightforward manner cannot lead to any upper bound better than cubic.

Instead, we perform an involved surgery on the path. The first idea is to start
with a path $\rho_\circ$ that is not the shortest, but uses the fewest zero
tests; the observation above shows that their number is bounded by $n$. Each
subpath between two consecutive zero tests is called an {\em{arc}}, and we aim
at modifying each arc separately to make it short. An arc is called {\em{low}}
if it contains only configurations with counter values at most $5n$, and
{\em{high}} otherwise. The total length of low arcs can again be bounded
by $O(n^2)$ by just excluding repeated configurations, so it suffices to focus
on high arcs.

Suppose $\rho$ is a high arc. Since we observe high counter values on $\rho$,
one can easily find a {\em{positive cycle}} $\sigma^+$ in the early parts
of $\rho$, and a {\em{negative cycle}} $\sigma^-$ in the late parts of $\rho$.
Here by a cycle we mean a sequence of transitions that starts and ends in the
same state, and the cycle is positive/negative if the total effect it has on the
counter during its traversal is positive/negative. Let $A$ be the (positive) effect
of $\sigma^+$ on the counter, and $-B$ be the (negative) effect of $\sigma^-$.

Now comes the crucial idea of the proof: we can modify $\rho$
by pumping $\sigma^+$ and $\sigma^-$ up many times, thus effectively ``lifting''
the central part of the path (called {\em{cap}}) to counter levels where there
is no threat of hitting counter value zero while performing modifications
(see Figure~\ref{fig:normal-arc}, p.~\pageref{fig:normal-arc}).
More importantly, the cap can now be unpumped
``modulo $\gcd(A,B)$'' in the following sense: we can exhaustively remove
subpaths between configurations that have the same state and whose counter
values are congruent modulo $\gcd(A,B)$. The reason is that any change in the
total effect of the cap on the counter that is divisible by $\gcd(A,B)$ can be
compensated by adjusting the number of times we pump cycles $\sigma^+$
and $\sigma^-$. In particular, the length of the cap becomes reduced
to at most $\gcd(A,B)\cdot n$, at the cost of pumping $\sigma^+$ and $\sigma^-$
several times.

By performing this operation (we call it \emph{normalization})
on all high arcs, we make them {\em{normal}}. After this, we apply an involved amortization
scheme to show that the total length of normal arcs is at most quadratic
in $n$. This requires very delicate arguments for bounding (\textit{i}) the total
length of the caps and (\textit{ii}) the total length of the pumped cycles $\sigma^+$
and $\sigma^-$ throughout all the normal arcs. In particular, for this part of the
proof to work we need to assert a number of technical properties of normal arcs;
we ensure that these properties hold when we perform the normalization.
Most importantly, whenever for two arcs the
corresponding cycles $\sigma^+$ (or $\sigma^-$) lie in the same strongly
connected component of the system (looking at the graph induced only by
non-zero transitions), we stipulate that in both arcs $\sigma^+$
(or $\sigma^-$) actually refer to the same cycle. The final amortization is based
on the analysis of pairs of strongly connected components to which $\sigma^+$
and $\sigma^-$ belong, for all normal arcs.

The way our proof modifies individual arcs extends the construction
found in Del\'eage and Pierre~\cite{DP86}.
In contrast to their work,
our treatment of automata with zero tests requires a global argument,
and for that a more refined modification of arcs and sophisticated
global analysis seem necessary.
At least as of now, arguments of this flavor (inspired by \emph{amortized
analysis} reasoning) are not typical for formal language
theory and are more characteristic of the body of work
on algorithms and data structures; see, e.g.,~\cite{Kozen-DAA,CormenLR-book}.


\section{Preliminaries}
\label{s:prelim}

In this paper $\N$ stands for the set of nonnegative integers.
For any set $X$ and a word $w \in X^*$, the {\em{length}} of $w = x_1 \ldots x_n$,
denoted $\len(w)$, is the number $n$ of symbols in $w$. For $k\in \N$ and
a word $w$, by $w^k$ we denote the word $w$ repeated $k$ times. For two
positive integers $x,y$, by $\gcd(x,y)$ and $\lcm(x,y)$ we denote the greatest
common divisor and the least common multiple of $x$ and~$y$, respectively.
Recall that $x\cdot y=\gcd(x,y)\cdot \lcm(x,y)$.

We now give all definitions related to one-counter systems
that we will need later.
For the reader's convenience, concepts from Section~\ref{s:overview}
are defined anew.

A~\emph{one-counter system} (OCS)~\OCS consists of a finite set of \emph{states} $Q$,
a set of \emph{non-zero transitions} $T_{>0} \subseteq Q \times \{-1,0,1\} \times Q$,
and a set of \emph{zero tests} $T_{=0} \subseteq Q \times \{0, 1\} \times Q$.
The set of \emph{transitions} is $T = T_{>0} \cup T_{=0}$. For a transition $t = (q, d, q') \in T$, by
$\source(t)$ and $\target(t)$
we denote $q$ and $q'$, i.e., the source and the target state of $t$
respectively. Further, the \emph{effect} of the transition $t = (q, d, q')$ is the number $d$; we write $\eff(t) = d$.
We extend this notion to sequences of transitions:
$\eff(t_1 \ldots t_m) = \sum_{i=1}^m \eff(t_i)$.

A \emph{configuration} of the OCS~\OCS is a pair in $Q \times \N$.
The \emph{state of a configuration} $(q, c)$ is the state $q$;
we also say that configuration $(q, c)$ \emph{has state} $q$, and write $\state((q, c)) = q$.
The \emph{counter value} of configuration $(q, c)$ is the number $c$;
we write $\cvalue((q, c)) = c$.

A transition $t = (q, d, q') \in T$ can be \emph{fired} in a configuration $\gamma = (q, c)$
if either $t \in T_{>0}$ and $c > 0$ or $t \in T_{=0}$ and $c = 0$.
In other words, zero tests can be fired only if the counter value
is zero, and non-zero transitions can be fired only if the counter value is positive.
The \emph{result} of firing $(q, d, q')$ in $(q, c)$ is the configuration $\gamma' = (q', c+d)$.
We then write $\gamma \trans{t} \gamma'$.

A \emph{path} $\rho$ of the OCS~\OCS is a sequence of pairs
\[
(\gamma_1, t_1) (\gamma_2, t_2) \ldots (\gamma_m, t_m) \in ((Q \times \N) \times T)^*
\]
such that for every $i \in \{1, \ldots, m-1\}$ we have $\gamma_i \trans{t_i} \gamma_{i+1}$
and there exists a configuration $\gamma_{m+1}$ such that $\gamma_m \trans{t_m} \gamma_{m+1}$.
The \emph{length} of this path is $m$.
The \emph{source} of $\rho$, denoted by $\source(\rho)$, is $\gamma_1$; we also say that $\rho$
\emph{starts} in its source.
Similarly, the \emph{target} of $\rho$, denoted by $\target(\rho)$, is $\gamma_{m+1}$; we say
that $\rho$ \emph{finishes} in its target. Note that 
now the source and target are configurations, rather than individual states;
the path is \emph{from} its source \emph{to} its target. 
All $\gamma_2, \ldots, \gamma_m$ are called \emph{intermediate configurations}.
We also say that configurations $\gamma_1,\gamma_2,\ldots,\gamma_{m+1}$
{\em{appear}} on $\rho$; note that the target of $\rho$ also appears on $\rho$.
Finally, when such a path exists, the configuration $\gamma_{m+1}$ is said to be
\emph{reachable} from the configuration $\gamma_1$.

The \emph{projection} of a path $\rho$ is the sequence of its transitions $t_1 t_2 \ldots t_m$;
we write $\ctproj(\rho) = t_1 t_2 \ldots t_m$.
We follow the convention of denoting paths by $\rho$ and sequences of transitions by $\sigma$.
The \emph{effect} of a path $\rho$ is $\eff(\rho)=\eff(\ctproj(\rho))$.
A sequence of transitions $\sigma = t_1 t_2 \dots t_m$ is \emph{fireable} in a configuration
$\gamma_1$ if there exists a path $\rho = (\gamma_1, t_1) (\gamma_2, t_2) \ldots (\gamma_m, t_m)$.
This path $\rho$ is called the \emph{fastening} of $\sigma$ at $\gamma_1$,
denoted $\rho = \fasten(\gamma_1, \sigma)$.
Note that in particular $\ctproj(\fasten(\gamma, \sigma)) = \sigma$
for every $\gamma$ in which $\sigma$ is fireable.

A sequence of transitions $t_1 t_2 \ldots t_m$ is \emph{consistent}
if for all $i \in \{1, \ldots, m-1\}$
it holds that $\target(t_i) = \source(t_{i+1})$.
Note that a sequence of transitions fireable in some
configuration is always consistent, but the other implication does not hold in
general. We extend the notation $\source(\cdot)$ and $\target(\cdot)$ to
consistent sequences of transitions: $\source(t_1 t_2 \ldots t_m)=\source(t_1)$
and $\target(t_1 t_2 \ldots t_m)=\target(t_m)$. The sources and targets of the
transitions of $t_1 t_2 \ldots t_m$ are {\em{visited}} on $t_1 t_2 \ldots t_m$.

A \emph{cycle} $\sigma$ is a consistent sequence of non-zero transitions that
starts and finishes in the same state $q$.
This $q$ is called the \emph{base state of the cycle} $\sigma$.
If the effect of $\sigma$ is positive (resp.\ negative), then it is
a \emph{positive} (resp. {\em{negative}}) cycle. A cycle $\sigma$ is called
{\em{simple}} if every state is visited at most once on $\sigma$, except for
the base of $\sigma$, which is visited only at the start and at the end.


\section{Proof of Theorem~\ref{thm:main-thm}}
\label{s:proof}

\subsection{Proof overview and notation}

Let us fix the OCS \OCS we work with; let $Q$ be its state set
and let $n=|Q|$. Suppose $\rho_0$ is a path from $\alpha$ to $\beta$,
and let $\rho_0$ be chosen such that it has the smallest possible number
of configurations with counter value zero.
Note that $\rho_0$ does not have to be the shortest path between $\alpha$ and $\beta$.
The first step is to divide $\rho_0$ into subpaths, called {\em{arcs}}, between
consecutive configurations with counter value zero.
Then we modify the arcs separately. If a counter value in an arc does not
exceed $5n$, then we say that the arc is \emph{low}, otherwise it is \emph{high}.
The low arcs will not be changed at all, and the reason is that we can bound
quadratically the total number of configurations with counter value
at most $5n$ using the following straightforward proposition.
It is similar, in the spirit, to pumping lemmas, but simply removes
a part of the path.

\begin{prop}\label{prop:basic-pumping}
Suppose $\rho=(\gamma_1, t_1) (\gamma_2, t_2) \ldots (\gamma_m, t_m)$ is a path
from $\alpha$ to $\beta$.
Suppose further that for some $i$ and $j$ with $1\leq i<j\leq m+1$ it holds
that $\gamma_i=\gamma_j$, where $\gamma_{m+1}$ is such
that $\gamma_m \trans{t_m} \gamma_{m+1}$.
Consider
$$\rho'=(\gamma_1, t_1) (\gamma_2, t_2)\ldots (\gamma_{i-1},t_{i-1})
(\gamma_j,t_j)
(\gamma_{j+1},t_{j+1})\ldots (\gamma_m,t_m).$$
Then $\rho'$ is also a path from $\alpha$ to $\beta$.
\end{prop}

However, the high arcs will be heavily modified.
Roughly speaking, if an arc is high, then it contains both a positive cycle
near its beginning and a negative cycle near its end. We can use these cycles to
pump the middle part of the path as much up as we like. Thus, the modified path
will consist of a short prefix;
then several iterations of a positive cycle pumping it up; then a so called {\em{cap}}: a part
of the path with only high counter values;
then several iterations of a negative cycle pumping it down; and finally a short suffix.
We show in the sequel how to perform this construction in such a way
that the total length of pumping cycles, short prefixes and suffixes, and caps is quadratic.
The construction itself (with arc-level length estimates)
is presented in the following Section~\ref{s:proof:normal},
and the upper bound on the length of the entire path
is given in Section~\ref{s:proof:global}.

\medskip
\subsubsection*{Transition multigraph.}
One can view a transition $(p, c, q) \in Q \times \{-1, 0, 1\} \times Q$
also as an edge $(p, q) \in Q \times Q$ labelled by a number $c \in \{-1, 0, 1\}$.
In the proof we will many times switch back and forth between these two perspectives.
In order to keep the mathematical precision we introduce a bit of notation.

The \emph{transition multigraph} $G = (V, E, \ell)$ of an OCS consists of a set of vertices~$V$,
a multiset of directed edges $E$, and a labeling $\ell: E \to \{-1, 0, 1\}$.
The set $V$ equals the set of states $Q$. Every 
non-zero transition $t = (p, c, q) \in T_{>0}$
in~\OCS gives rise to an edge $e = (u, v) \in E$ with $\ell(e) = c$.
Note that the definition of the transition multigraph does not take into
account zero transitions.

In the proof we pay special attention to strongly connected components (SCCs)
of $G$.
Recall that two vertices $p,q\in V$ are said to {\em{communicate}} if $G$ has a walk
from $p$ to $q$ and a walk from $q$ to $p$. Communication is an equivalence
relation, and its equivalence classes are called the {\em{strongly connected
components}} of $G$. Let $\SCCs$ be the set of all strongly connected
components of $G$. For a strongly connected component $S\in \SCCs$, by $n_S$ we
denote the number of vertices in $S$. We say that a cycle $\sigma$ is
{\em{contained}} in $S$ if each state appearing on $\sigma$ belongs to $S$.
Note that every cycle is contained in some SCC,
and a simple cycle contained in $S$ has length at most $n_S$.
We say that an SCC $S$ is \emph{positively enabled} if it contains
a cycle that has a positive effect.
Similarly, $S$ is \emph{negatively enabled} if it contains a cycle that has a negative effect.
Note that an SCC $S$ can be both positively and negatively enabled.

\begin{lem}\label{lem:short-positive-cycle}
Let $G$ be a transition multigraph of an OCS
and $S$ a positively (respectively, negatively) enabled SCC.
Then there exists a positive (respectively, negative)
cycle $\sigma$ contained in $S$ that is simple.
\end{lem}

\begin{proof}
We prove the lemma for positively enabled SCCs; the proof for negatively
enabled SCCs is symmetric.
By definition, $S$ contains a positive cycle $\sigma$.
Choose $\sigma$ to be the shortest such cycle; we claim that then
$\sigma$ is simple. Aiming towards a contradiction, suppose that some state
repeats on $\sigma$.
Then $\sigma$ can be decomposed into two cycles $\sigma_1,\sigma_2$
that are strictly shorter than $\sigma$. Since $\sigma$ is
positive and $\eff(\sigma)=\eff(\sigma_1)+\eff(\sigma_2)$,
we infer that either $\sigma_1$ or $\sigma_2$ is positive. This
contradicts the minimality of $\sigma$.
\end{proof}

For every positively enabled SCC $S$ we distinguish one,
arbitrarily chosen, simple cycle with positive effect
contained in $S$; we denote it by $\sigma^{+}_S$.
Its existence is guaranteed by Lemma~\ref{lem:short-positive-cycle}.
Similarly, for every negatively enabled $S$
we distinguish one simple cycle with negative effect contained in $S$,
and we denote it by $\sigma^{-}_S$.
The base states of these cycles are chosen arbitrarily.

\subsection{Normal paths}
\label{s:proof:normal}

A path is an \emph{arc} if both its source and target have counter value zero,
but all its intermediate configurations
have counter values strictly larger than zero.
An arc (or a path) is \emph{low} if all its configurations (including the target)
have counter values strictly smaller than $5n$.
An arc $\rho$ is \emph{$(S, T)$-normal},
where $S$ and $T$ are some SCCs of the transition multigraph,
if it admits the following \emph{normal decomposition}
(see Figure~\ref{fig:normal-arc}, p.~\pageref{fig:normal-arc}):
\[
\rho = \rho_\pref \, \rho_\up \, \rho_\rcap \, \rho_\down \, \rho_\suff,
\]
where
\begin{itemize}[label=\textrm{--}]
  \item $\rho_\pref$ and $\rho_\suff$ are low;
  \item $\ctproj(\rho_\up) = (\sigma_\up)^k$ for some $k \in \N$,
  where $\sigma_\up = \sigma^{+}_S$;
  \item  $\ctproj(\rho_\down) = (\sigma_\down)^\ell$ for some $\ell \in \N$,
  where $\sigma_\down = \sigma^{-}_T$; 
  \item $\state(\source(\rho_\rcap))$ is the base state of $\sigma_\up$; and
  \item $\state(\target(\rho_\rcap))$ is the base state of $\sigma_\down$.
\end{itemize}
We say that an arc $\rho$ is {\em{normal}} if it is $(S,T)$-normal
for some $S,T\in \SCCs$.
Then a path $\rho'$ is \emph{normal} if it is a concatenation
of normal arcs (possibly for different pairs $(S,T)$) and low arcs.

In the remaining part of the proof we will show
that if $\beta$ is reachable from $\alpha$, where
$\cvalue(\alpha) = \cvalue(\beta) = 0$,
then there exists a short normal path from $\alpha$ to $\beta$.
We start by analyzing a single arc.

The following lemma, which is the most technically involved step in this paper,
shows that we can restrict ourselves to normal arcs that have a very special structure.
The proof of the lemma relies on ideas already present in Del\'eage and
Pierre~\cite{DP86}; we need, however, a much more refined statement for the
subsequent (``global'') part of our proof (Section~\ref{s:proof:global}).

\begin{lem}\label{lem:normalization}
If $\cvalue(\alpha) = \cvalue(\beta) = 0$ and there exists an arc from $\alpha$ to $\beta$,
then there exists an arc $\rho$ from $\alpha$ to $\beta$ which is either low or normal.
Moreover, in the case when $\rho$ is normal, a normal decomposition
$\rho = \rho_\pref \, \rho_\up \, \rho_\rcap \, \rho_\down \, \rho_\suff$
can be chosen such that:
\begin{enumerate}[label=\textup{(}\textit{\roman*}\textup{)}]
  \item\label{pr:up} $\ctproj(\rho_\up) = (\sigma_\up)^a$, $\eff(\sigma_\up) = A$ for some $a, A \in \N$;
  \item\label{pr:down} $\ctproj(\rho_\down) = (\sigma_\down)^b$, $\eff(\sigma_\down) = -B$ for some $b, B \in \N$;
  \item\label{pr:upl} $a \cdot A \leq 2\cdot\len(\rho_\rcap) + 2\cdot \lcm(A, B)$;
  \item\label{pr:downl} $b \cdot B \leq 2\cdot\len(\rho_\rcap) + 2\cdot \lcm(A, B)$;
  \item\label{pr:cap} no infix of $\ctproj(\rho_\rcap)$ is a cycle with effect divisible by $\gcd(A, B)$; 
  \item\label{pr:prsu-count} $\cvalue(\target(\rho_\up)),\cvalue(\source(\rho_\down))>n$; and
  \item\label{pr:prsu} all configurations appearing on $\rho_\pref$ and $\rho_\suff$ are pairwise different.
\end{enumerate}
\end{lem}

\noindent
We now explain some intuition behind this statement.
First note that, by condition~\ref{pr:prsu}, the total number of configurations
appearing on $\rho_\pref$ and $\rho_\suff$ is at most $5 n \cdot n$,
since $n$ is the number of states of the OCS~\OCS
and both of these paths are low (so counter values $5 n$ and above do not occur).
Thus, $\len(\rho_\pref) + \len(\rho_\suff) \le 5 n^2$.
Second, we can conclude from condition~\ref{pr:cap} that
every state $q \in Q$ can occur in configurations appearing in $\rho_\rcap$
at most $\gcd(A, B)$ times; hence, $\len(\rho_\rcap) \le n \cdot \gcd(A, B) \le n^2$.
Finally, condition~\ref{pr:up} implies $\len(\rho_\up) \le a \cdot n$;
if, for instance, $a \leq \const \cdot n$, then
$\len(\rho_\up) \le \const \cdot n^2$; similarly, $\len(\rho_\down) \le \const \cdot n^2$.
Combined together, these bounds would in this case show that $\len(\rho)$ is at most quadratic in~$n$.

However, this reasoning would be insufficient for our purposes, since
the number of normal arcs itself can be linear in~$n$.
This motivates more subtle upper bounds~\ref{pr:upl} and~\ref{pr:downl}
and the fine-grained choice of parameter in~\ref{pr:cap}.
We show how to use Lemma~\ref{lem:normalization} to obtain a quadratic upper bound
on the size of the \emph{entire} path in the following Section~\ref{s:proof:global};
the remainder of the present Section proves Lemma~\ref{lem:normalization}.

\begin{proof}
Fix configurations $\alpha$ and $\beta$ such
that $\cvalue(\alpha) = \cvalue(\beta) = 0$
and there exists an arc from $\alpha$ to $\beta$.
If there is a low arc from $\alpha$ to $\beta$, then there is nothing to prove,
so assume that all the arcs from $\alpha$ to $\beta$ are not low. Let $\rho_\circ$
be such an arc of the shortest possible length; then $\rho_\circ$ is not low. Let
\[
\rho_\circ = (\gamma_1, t_1) \ldots (\gamma_m, t_m),
\]
where $\alpha = \gamma_1$ and $\gamma_m \trans{t_m} \gamma_{m+1} = \beta$.
Since $\rho_\circ$ is shortest possible, from Proposition~\ref{prop:basic-pumping}
we infer that configurations $\gamma_1,\gamma_2,\ldots,\gamma_{m+1}$ are
pairwise different.

\begin{figure}[t]
                \centering
                \def\svgwidth{0.8\columnwidth}
                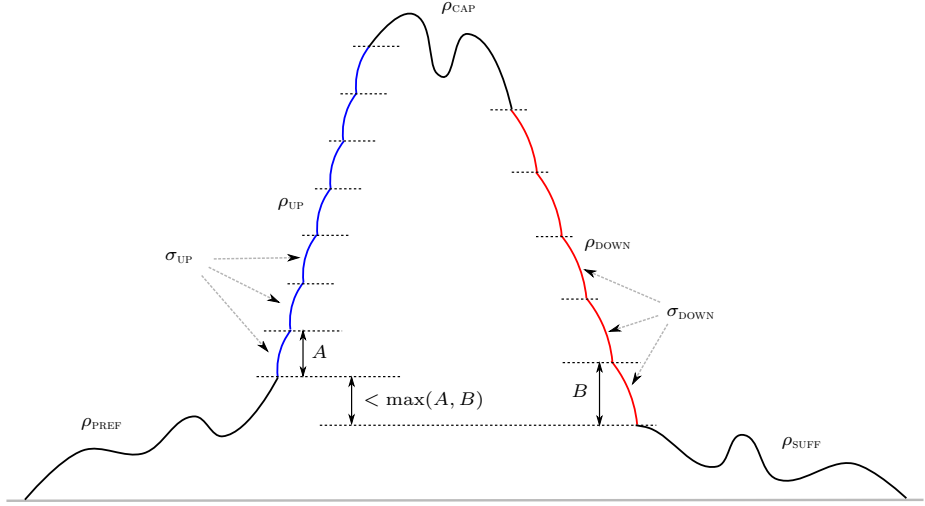
\caption{The normal decomposition of an arc together with some notation from
the proof of Lemma~\ref{lem:normalization}.}\label{fig:normal-arc}
\end{figure}

We start with a short overview.
Based on $\rho_\circ$ we construct a normal arc $\rho$ from $\alpha$ to $\beta$
satisfying the promised conditions.
Roughly speaking we proceed as follows.
First, we carefully define $\rho_\pref$ and $\rho_\suff$ so that
condition~\ref{pr:prsu} is satisfied; in this step we also fix
the components $S,T\in \SCCs$ for which $\rho$ will be $(S,T)$-normal.
Then we construct a sequence of transitions $\sigma_\rcap$ that, after
fastening it at some configuration, will form $\rho_\rcap$ that satisfies
condition~\ref{pr:cap}. Intuitively, $\sigma_\rcap$ is formed by exhaustively
unpumping the middle part of $\rho_\circ$. As $S,T$ are already fixed, so are also
cycles $\sigma_\up=\sigma^+_S$ and $\sigma_\down=\sigma^-_T$. Hence at this
point to completely define $\rho$ it remains to choose numbers $a$ and $b$.
At the end we show that $a,b$ can be chosen so that $\rho$ is indeed a valid path,
and moreover conditions~\ref{pr:upl} and~\ref{pr:downl} are satisfied.

Let us carry out this plan.
Consider any $k$ with $2n \leq k \leq 3n$. 
Let $i_k$ be the smallest index for which $\cvalue(\gamma_{i_k})=k$,
and let $j_k$ be the largest index for which $\cvalue(\gamma_{j_k})=k$.
Clearly such configurations exist, because $\rho_\circ$ is not low.
It moreover holds that $i_{2n} < i_{2n+1} < \ldots < i_{3n} < j_{3n} < \ldots < j_{2n+1} < j_{2n}$.
By the pigeonhole principle there exist indices $k, \ell$, where $k<\ell$, such that
the state of $\gamma_{i_k}$ equals the state of $\gamma_{i_\ell}$. Let this state be $p \in Q$.
\label{text:state-p}
Consider the sequence of transitions
\[
\sigma_\cyc = t_{i_k}t_{i_k+1} \ldots t_{i_\ell - 1}.
\]
It follows that $\sigma_\cyc$ is a positive cycle
with effect $\ell - k$ and base state $p$.
Let $S$ be the SCC of $G$ in which $\sigma_\cyc$ is contained;
the existence of $\sigma_\cyc$ asserts that $S$ is positively enabled.
Let $\sigma_\up = \sigma^{+}_S$.

Let $\tilde{\rho}_\pref$ be the prefix of $\rho_\circ$
up to configuration $\gamma_{i_k}$
(i.e., with $\target(\tilde{\rho}_\pref)=\gamma_{i_k}$).
Note that we cannot simply put $\rho_\pref=\tilde{\rho}_\pref$, because the
state $p$ in which $\tilde{\rho}_\pref$ finishes does not have to be the base
state of $\sigma_\up$, which
is the cycle that is required to be the one used for constructing $\rho_\up$.
This, however, poses no real difficulty, because $p$ and $\sigma_\up$ are
contained in the same SCC $S$, so we can easily augment $\tilde{\rho}_\pref$ by
a path to the base state of $\sigma_\up$ as follows.

Precisely we do the following. Let $q$ be the base state of $\sigma_\up$. As
both $p$ and $q$ belong to $S$, there exist consistent sequences of non-zero
transitions $\sigma_{pq}$ and $\sigma_{qp}$, leading from $p$ to $q$ and
from $q$ to $p$, respectively, such that:
\begin{enumerate}[label=(\alph*)]
  \item\label{pr2:short} states visited on $\sigma_{pq}$ are pairwise
                         different, and the same holds also for $\sigma_{qp}$;
                         in particular $\len(\sigma_{pq}), \len(\sigma_{qp}) < n_S$;
  \item\label{pr2:pq} $\sigma_{pq}$ is fireable
                      in any configuration $(p, c)$ for any $c \geq n$; and
  \item\label{pr2:qp} $\sigma_{qp}$ is fireable
                      in any configuration $(q, c)$ for any $c \geq n$.
\end{enumerate}
Assertion~\ref{pr2:short} follows from the fact that $\sigma_{pq}$
and $\sigma_{qp}$ can be chosen so that they correspond to simple paths in $G$,
i.e., walks with no state repeated.
Assertion~\ref{pr2:short} in particular implies that
the effects of prefixes of $\sigma_{pq}$ and $\sigma_{qp}$ are strictly larger than $-n$.
This implies assertions~\ref{pr2:pq} and~\ref{pr2:qp}.

Now we construct a path $\rho''_\pref$ as follows. Let $\rho_{pq}$ be the
fastening of $\sigma_{pq}$ at the configuration $\gamma_{i_k}$.
The state of $\gamma_{i_k}$ is $p$ and its counter value is not smaller
than $2n$, so indeed $\sigma_{pq}$ is fireable from $\gamma_{i_k}$; even more,
since $\len(\sigma_{pq})<n$ and $\cvalue(\gamma_{i_k})\geq 2n$, all the counter
values on $\rho_{pq}$ are larger than $n$. We define then
\[
\rho''_\pref = \tilde{\rho}_\pref \rho_{pq}=
(\gamma_1, t_1) \cdots (\gamma_{i_k - 1}, t_{i_k - 1}) \, \rho_{pq}.
\]

We construct path $\rho''_\suff$ in a completely symmetric manner, so we only
make a short summary in order to introduce the notation.
By the pigeonhole principle, for some $\bar{\ell}, \bar{k}$ with $\bar{\ell} < \bar{k}$
the state of $\gamma_{j_{\bar{\ell}}}$ and $\gamma_{j_{\bar{k}}}$ is the same,
let it be $\bar{p}$. \label{text:state-bar-p}
The part of the path between $\gamma_{j_{\bar{\ell}}}$
and $\gamma_{j_{\bar{k}}}$ projects to a negative cycle, so it is contained in some
negatively enabled SCC $T$, to which $\bar{p}$ also belongs.
Define $\sigma_\down = \sigma^{-}_{T}$, and let $\bar{q}$
be the base state of $\sigma_\down$.
As $\bar{p}$ and $\bar{q}$ both belong to $T$,
we have $\sigma_{\bar{q}\bar{p}}$ and $\sigma_{\bar{p}\bar{q}}$
with similar properties as $\sigma_{pq}$ and $\sigma_{qp}$.
Path $\rho_{\bar{q}\bar{p}}$ can be again defined as an appropriate fastening
of $\sigma_{\bar{q}\bar{p}}$, so we define
\[
\rho''_\suff =
\rho_{\bar{q}\bar{p}} \, (\gamma_{j_{\bar{k}}}, t_{j_{\bar{k}}}) \cdots (\gamma_m, t_m).
\]

Let $A = \eff(\sigma_\up)$ and $B = -\eff(\sigma_\down)$.
Now, based on $\rho''_\pref$ and $\rho''_\suff$ we define $\rho'_\pref$ and
$\rho'_\suff$ as follows.
Observe that $\cvalue(\target(\rho''_\pref))=k+\eff(\sigma_{pq})>2n-n=n$, and
similarly $\cvalue(\source(\rho''_\suff))>n$.
Suppose first that
$\cvalue(\target(\rho''_\pref))\leq \cvalue(\source(\rho''_\suff))-A$.
Then we take $\rho'_\suff=\rho''_\suff$,
whereas $\rho'_\pref$ is obtained from $\rho''_\pref$ by appending
the cycle $\sigma_\up$ a number of times so that
$\cvalue(\source(\rho'_\suff))-A<\cvalue(\target(\rho'_\pref{)})
\leq \cvalue(\source(\rho'_\suff))$.
Similarly, if $\cvalue(\target(\rho''_\pref))\geq \cvalue(\source(\rho''_\suff))+B$,
then we take $\rho'_\pref=\rho''_\pref$ whereas $\rho'_\suff$
is constructed from $\rho''_\suff$ by appending $\sigma_\down$
a number of times in the front so that
$\cvalue(\target(\rho'_\pref))-B<\cvalue(\source(\rho'_\suff))
\leq \cvalue(\target(\rho'_\pref))$.
If none of these cases holds, we simply take $\rho'_\pref=\rho''_\pref$
and $\rho'_\suff=\rho''_\suff$.
Since $\sigma_\up, \sigma_\down$ have lengths at most $n$,
and the first one is a positive cycle whereas the second one is a negative
cycle, it can be easily verified that $\rho'_\pref$ and $\rho'_\suff$ are
indeed valid paths; here we use the property that
$\cvalue(\target(\rho''_\pref))>n$ and $\cvalue(\source(\rho''_\suff))>n$
in order to make sure that appending the cycles does not create nonpositive
counter values on the path.
Moreover, we achieved the property that
$|\cvalue(\target(\rho'_\pref))-\cvalue(\source(\rho'_\suff))|<\max(A,B)$.

Finally, we obtain $\rho_\pref$ by applying
Proposition~\ref{prop:basic-pumping} to $\rho'_\pref$ exhaustively. In this
manner $\rho_\pref$ has still the same source and target as $\rho'_\pref$, but
no configuration repeats on $\rho_\pref$. Similarly, $\rho_\suff$ is obtained
from $\rho'_\suff$ by applying Proposition~\ref{prop:basic-pumping}
exhaustively, so that no configuration repeats on $\rho_\suff$.

Let $\zeta=\target(\rho_\pref)=\target(\rho'_\pref)$
and $\bar{\zeta}=\source(\rho_\suff)=\source(\rho'_\suff)$.
We now verify that $\rho_\pref$ and $\rho_\suff$ are as required.

\begin{clm}\label{cl:prefsuff}
The following conditions hold:
\begin{enumerate}[label=(\alph*)]
\item\label{ps:low} paths $\rho_\pref$ and $\rho_\suff$ are low;
\item\label{ps:non-zero} the counter values in configurations
    appearing on $\rho_\pref$ and $\rho_\suff$ are always positive,
    apart from the source of $\rho_\pref$ (which is $\alpha$)
    and the target of $\rho_\suff$ (which is $\beta$);
\item\label{ps:counter} $\cvalue(\zeta),\cvalue(\bar{\zeta})>n$;
\item\label{ps:diff} $|\eff(\rho_\pref)+\eff(\rho_\suff)|<\max(A,B)$; 
\item\label{ps:prop} property~\ref{pr:prsu} is satisfied.
\end{enumerate}
\end{clm}

\begin{proof}
By the definition of $k$, all the configurations appearing
on $\tilde{\rho}_\pref$ have counter values at most $3n$.
Since the counter value of configuration $\gamma_{i_k}$ is not smaller that $2n$
and not larger than $3n$, and $|\eff(\sigma_{pq})|\leq \len(\sigma_{pq})<n$,
we infer that the counter value on the path $\rho''_\pref$
is always strictly smaller than $4n$. As $\len(\sigma_\up)<n$, it can be easily
seen that appending the cycles during the construction of $\rho'_\pref$ cannot
create counter values larger than $5n-1$. Hence $\rho'_\pref$ is low, and
consequently $\rho_\pref$ is also low. A symmetric reasoning shows the same
conclusions for $\rho_\suff$, and thus condition \ref{cl:prefsuff}\ref{ps:low}
is satisfied.

Since $\rho_\circ$ is an arc, no configuration on $\tilde{\rho}_\pref$
apart from $\alpha$ has nonpositive counter value.
As $\cvalue(\gamma_{i_k})\geq 2n$, we have already argued that both after
adding $\sigma_{pq}$ when constructing $\rho''_\pref$,
and after adding cycles $\sigma_\up$ when constructing $\rho'_\pref$, we could
not obtain a configuration with a nonpositive counter value. Hence the only
configuration on $\rho'_\pref$ that has zero counter value is $\alpha$, and the
same holds also for $\rho_\pref$. A symmetric reasoning yields a symmetric
conclusion for $\rho_\suff$, which proves condition~\ref{cl:prefsuff}\ref{ps:non-zero}.

From the construction we have $\cvalue(\zeta)\geq \cvalue(\target(\rho''_\pref))$,
and we already argued that $\cvalue(\target(\rho''_\pref))>n$.
Hence $\cvalue(\zeta)>n$,
and a symmetric reasoning shows that $\cvalue(\bar{\zeta})>n$. This proves
condition~\ref{cl:prefsuff}\ref{ps:counter}.

Observe that $\eff(\rho_\pref)=\cvalue(\zeta)$
and $\eff(\rho_\suff)=-\cvalue(\bar{\zeta})$.
Hence, condition~\ref{cl:prefsuff}\ref{ps:diff}
follows from $|\cvalue(\zeta)-\cvalue(\bar{\zeta})|<\max(A,B)$.

For condition~\ref{cl:prefsuff}\ref{ps:prop},
aiming towards a contradiction suppose that some configuration $\gamma$ appears
more than once on $\rho_\pref$ and $\rho_\suff$. By construction, no
configuration repeats on $\rho_\pref$ and on $\rho_\suff$ individually, so one
of the appearances of $\gamma$ is on $\rho_\pref$ and the second is on $\rho_\suff$.
Define a path $\rho_1$ by concatenating the prefix of $\rho_\pref$
up to the appearance of $\gamma$ together with the suffix of $\rho_\suff$ beginning from
the appearance of $\gamma$. Clearly, $\rho_1$ is an arc from $\alpha$ to $\beta$,
and moreover it is low because both $\rho_\pref$ and $\rho_\suff$ are
low. This contradicts the assumption that there is no low arc from $\alpha$ to $\beta$.\end{proof}

The intuition now is that by repeating $\sigma_\up$ and $\sigma_\down$
appropriately many times (i.e., selecting numbers $a$ and $b$) we can choose
any difference of effects of $\rho_\pref\rho_\up$ and $\rho_\down\rho_\suff$,
as long as this difference belongs to a fixed congruence class modulo $\gcd(A,B)$.
This means that the middle part of the path $\rho_\circ$ can be
unpumped ``modulo $\gcd(A,B)$'': even if we change its effect by a multiple
of $\gcd(A,B)$, we will be able to compensate for this change by adjusting $a$
and $b$.

We now proceed to showing how the middle part of the path, i.e., $\rho_\rcap$,
will be constructed. Intuitively, the idea is to take the part from $\rho_\circ$
between indices $k$ and $\bar{k}$, augment it with short connectives $\sigma_{qp}$
and $\sigma_{\bar{p}\bar{q}}$ to link it with the cycles $\sigma_\up$
and $\sigma_\down$, and unpump it ``modulo $\gcd(A,B)$''
exhaustively. However, during further constructions we need certain
divisibility properties of $\eff(\sigma_\rcap)$, and hence the construction of
the connections to $\sigma_\up$ and $\sigma_\down$ is more complicated.

\begin{clm}\label{cl:good-cap}
There exists a sequence of transitions $\sigma_\rcap$ such that
\begin{enumerate}[label=(\alph*)]
  \item\label{cap:stfi} $\sigma_\rcap$ starts in $q$ and finishes in $\bar{q}$ 
 (base states of $\sigma_\up$ and $\sigma_\down$ respectively);
  \item\label{cap:gcd} no infix of $\sigma_\rcap$ is a cycle
                       with effect divisible by $\gcd(A, B)$; and
  \item\label{cap:sum} $\eff(\rho_\pref) + \eff(\sigma_\rcap) + \eff(\rho_\suff)$
                       is divisible by $\gcd(A, B)$.
\end{enumerate}
\end{clm}

\begin{proof}
The construction is depicted in Figure~\ref{fig:pumping}. Let $\sigma_\preconn = \sigma_{pq} \sigma_{qp}$
and $\sigma_\postconn = \sigma_{\bar{p}\bar{q}} \sigma_{\bar{q}\bar{p}}$.
We set
\begin{align*}
\sigma_\premidd = \sigma_{qp} \, (\sigma_\preconn)^c\\
\sigma_\postmidd = (\sigma_\postconn)^c \, \sigma_{\bar{p}\bar{q}},
\end{align*}
where $c = \gcd(A, B) - 1$.
It is easy to verify that $\sigma_\premidd,\sigma_\postmidd$ are consistent and
\begin{align}
\eff(\sigma_\premidd) + \eff(\sigma_{pq}) =
(c+1) \cdot \eff(\sigma_\preconn) \equiv 0 \mod \gcd(A, B),\label{eq:pre}\\
\eff(\sigma_\postmidd) + \eff(\sigma_{\bar{q}\bar{p}}) =
(c+1) \cdot \eff(\sigma_\postconn) \equiv 0 \mod \gcd(A, B).\label{eq:post}
\end{align}
Recall that the overall arc
$
\rho_\circ = (\gamma_1, t_1) \ldots (\gamma_m, t_m)
$
visits the states $p$ and $\bar p$
(which were defined on pages~\pageref{text:state-p}--\pageref{text:state-bar-p});
the sequence of transitions
\[
\sigma_\midd = t_{i_k} t_{i_k + 1} \cdots t_{j_{\bar{k}}-1}
\]
is the fragment between these two visits
(more precisely, $\sigma_\midd$ can be obtained by removing
appropriate prefix and suffix from the sequence of transitions of $\rho_\circ$).
Define $\sigma'_\rcap=\sigma_\premidd \sigma_\midd \sigma_\postmidd$.
We now verify that conditions~\ref{cl:good-cap}\ref{cap:stfi}
and~\ref{cl:good-cap}\ref{cap:sum} are satisfied for $\sigma'_\rcap$.
Condition~\ref{cl:good-cap}\ref{cap:stfi} follows directly from the
construction. For condition~\ref{cl:good-cap}\ref{cap:sum}, observe that
from the construction of $\rho_\pref$ and $\rho_\suff$ we have
\begin{align*}
&\eff(\rho_\pref)=\eff(\rho'_\pref)=
\eff(t_1t_2\ldots t_{i_k-1})+\eff(\sigma_{pq})+x\cdot A\\
&\eff(\rho_\suff)=\eff(\rho'_\suff)=
\eff(t_{j_{\bar{k}}}t_{j_{\bar{k}}+1}\ldots t_{m})+\eff(\sigma_{\bar{q}\bar{p}})-y\cdot B
\end{align*}
for some $x,y\in \N$.
Hence, from~\eqref{eq:pre},~\eqref{eq:post} and the fact that $\eff(\rho_\circ)=0$,
it follows that
\begin{align*}
\eff(\rho_\pref) + \eff(\sigma'_\rcap) + \eff(\rho_\suff) & \equiv\\
\eff(\rho_\circ)+\eff(\sigma_{pq})+\eff(\sigma_{\bar{q}\bar{p}})
+\eff(\sigma_\premidd)+\eff(\sigma_\postmidd) & \equiv 0 \mod \gcd(A,B).
\end{align*}
So condition~\ref{cl:good-cap}\ref{cap:sum} indeed holds for $\sigma'_\rcap$.

\begin{figure}[t]
                \centering
                \def\svgwidth{0.5\columnwidth}
                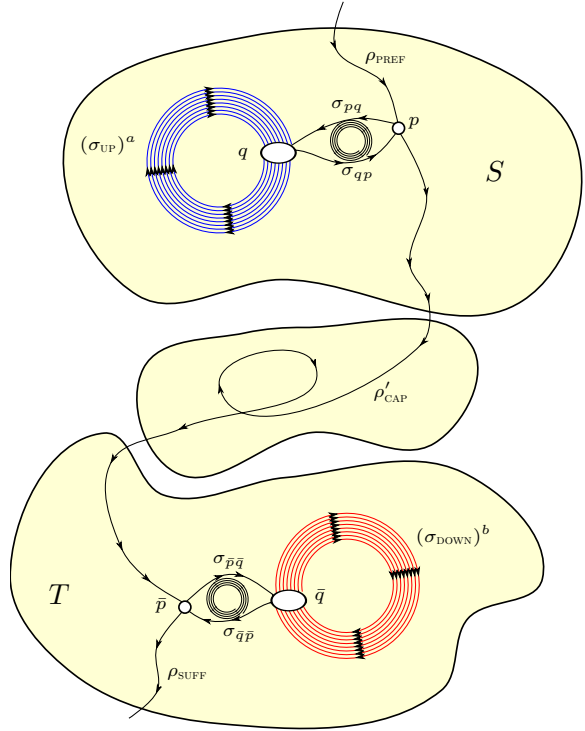
\caption{The construction of $\rho'_\rcap$ in the first part of the proof of Claim~\ref{cl:good-cap}. Note that cycles $(\sigma_\preconn)^c$ and $(\sigma_\postconn)^c$ are depicted symbolically as small spirals.}\label{fig:pumping}
\end{figure}

We now take condition~\ref{cl:good-cap}\ref{cap:gcd} into consideration.
Define $\sigma_\rcap$ to be any of the shortest possible consistent sequences of
transitions satisfying conditions~\ref{cl:good-cap}\ref{cap:stfi}
and~\ref{cl:good-cap}\ref{cap:sum}; the existence of $\sigma'_\rcap$ implies
that there exists such a sequence, so $\sigma_\rcap$ is well-defined.
Let $\sigma_\rcap = t_1 \ldots t_r$. Assume towards contradiction that $\sigma_\rcap$
does not satisfy condition~\ref{cl:good-cap}\ref{cap:gcd}.
Then there is some infix $t_{i+1} \ldots t_j$ that is a cycle and its effect
is divisible by $\gcd(A, B)$.
It is easy to observe that
sequence $\sigma' = t_1 \cdots t_i t_{j+1} \cdots t_m$ is consistent,
$\source(\sigma') = q$, $\target(\sigma') = \bar{q}$ and
$\eff(\sigma') - \eff(\sigma) \equiv 0 \mod \gcd(A, B)$.
Hence $\sigma'$ satisfies conditions~\ref{cl:good-cap}\ref{cap:stfi}
and~\ref{cl:good-cap}\ref{cap:sum} while being strictly shorter than
$\sigma_\rcap$. This contradicts the minimality (shortness) of $\sigma_\rcap$
and proves that $\sigma_\rcap$ satisfies
condition~\ref{cl:good-cap}\ref{cap:gcd}.\end{proof}

Note that from condition~\ref{cl:good-cap}\ref{cap:gcd} it follows that
$\len(\sigma_\rcap)\leq \gcd(A,B)\cdot n$. In the final construction this
condition will directly imply that $\rho_\rcap$ will satisfy
property~\ref{pr:cap}, since $\rho_\rcap$ will be simply $\sigma_\rcap$
fastened at some configuration.

We denote
\begin{align*}
L & = \len(\sigma_\rcap),\\
K & = \eff(\rho_\pref) + \eff(\sigma_\rcap) + \eff(\rho_\suff).
\end{align*}
Recall that $\gcd(A,B)$ divides $K$.
Moreover, from condition~\ref{cl:prefsuff}\ref{ps:diff} we know that 
\begin{equation}\label{eq:boundK}
|K|\leq |\eff(\sigma_\rcap)|+|\eff(\rho_\pref)+\eff(\rho_\suff)|<L+\max(A,B).
\end{equation}

Having defined $\rho_\pref$, $\rho_\suff$ and $\sigma_\rcap$, we proceed to
defining $\rho_\up$ and $\rho_\down$. For this, we need to define $a,b\in \N$:
the numbers of times the cycles $\sigma_\up$ and $\sigma_\down$ are repeated
on $\rho_\up$ and $\rho_\down$. As described earlier, they have to be chosen so
that the resulting path $\rho$ is valid and has zero effect, but they also need
to be reasonably small so that conditions~\ref{pr:upl} and~\ref{pr:downl}
are satisfied. We now prove that this is always possible.

\begin{clm}\label{cl:ab}
There exist $a,b\in \N$ such that the following conditions hold:
\begin{enumerate}[label=(\alph*)]
\item\label{ab:sum} $a\cdot A-b\cdot B = -K$;
\item\label{ab:inq} $L\leq a \cdot A, b \cdot B \leq 2L + 2\cdot \lcm(A, B)$.
\end{enumerate}
\end{clm}

\begin{proof} B\'{e}zout's identity states that
there exist some integers $x_0,y_0$
such that $x_0\cdot A-y_0\cdot B=\gcd(A,B)$.
Since $K$ is divisible by $\gcd(A,B)$, we can take $x=x_0\cdot (-K/\gcd(A,B))$
and $y=y_0\cdot (-K/\gcd(A,B))$ so that $x\cdot A-y\cdot B=-K$. Moreover, by
increasing $x$ by $M\cdot B$ and
increasing $y$ by $M\cdot A$, for a sufficiently large integer $M$, we can
further assume that $x,y\geq 0$. Suppose then that $(x,y)$ is a pair of
nonnegative integers satisfying $x\cdot A-y\cdot B=-K$ for which $x+y$ is the
smallest possible. We claim that $x \cdot A,y\cdot B \leq |K| + \lcm(A, B)$.

Aiming towards a contradiction, suppose that $x\cdot A>|K| + \lcm(A, B)$.
Then in particular $x>\lcm(A,B)/A=B/\gcd(A,B)$.
Also, $y\cdot B=x\cdot A+K>\lcm(A,B)$, and hence $y>\lcm(A,B)/B=A/\gcd(A,B)$.
Consider $x'=x-B/\gcd(A,B)$ and $y=y'-A/\gcd(A,B)$.
Then we have $x',y'\geq 0$ and it can be easily verified that
$x'\cdot A-y'\cdot B=-K$.
As $x'+y'<x+y$, this contradicts the minimality of $x+y$. Hence
indeed $x\cdot A\leq |K| + \lcm(A, B)$, and symmetrically we also prove that
$y\cdot B\leq |K| + \lcm(A, B)$. Note that by~\eqref{eq:boundK} the inequalities
$|K|+\lcm(A,B)\leq L+\max(A,B)+\lcm(A,B)\leq L+2\cdot \lcm(A,B)$ hold.

It remains to define $a,b$ based on $x,y$ so that the lower bound on $a\cdot A$
and $b\cdot B$ holds. If already $x\cdot A,y\cdot B\geq L$, then we can take
$(a,b)=(x,y)$; assume therefore that this is not the case. For $i\in \N$, let
\begin{equation*}
a_i = x + i \cdot (\lcm(A, B)/A)\quad\textrm{and}\quad b_i = y + i \cdot (\lcm(A, B)/B).
\end{equation*}
Clearly $a_i,b_i\geq 0$, and it is easy to verify that
$a_i\cdot A-b_i\cdot B=-K$ for each $i\in \N$.
It therefore only suffices to show that there exists $i$ such that
$L \leq a_i \cdot A, b_i \cdot B \leq 2L + 2\cdot\lcm(A, B)$.
Let $i$ be the smallest nonnegative integer so that $a_i\cdot A\geq L$
and $b_i \cdot B \geq L$; then $i>0$.
Suppose that $K\geq 0$; the other case is symmetric.
As $a_i\cdot A-b_i\cdot B=-K\leq 0$, we have $a_i\cdot A\leq b_i\cdot B$;
by our definition of $a_i$ and $b_i$, similarly
$a_{i-1} \cdot A \leq b_{i-1} \cdot B$.
Since by the minimality of $i$
either $a_{i-1} \cdot A < L$ or $b_{i-1} \cdot B < L$ holds,
we deduce that $a_{i-1} \cdot A < L$ in both cases,
and it follows that $a_i \cdot A < L + \lcm(A, B)$.
Now
\begin{equation*}
    b_i \cdot B
= K + a_i \cdot A
  < (L + \max(A, B)) + (L + \lcm(A, B))
\le 2 L + 2 \cdot \lcm(A, B).
\end{equation*}
Thus we can take $(a, b) = (a_i, b_i)$. The case $K < 0$ is symmetric.\end{proof}

Let us fix the numbers $a, b \in \N$ given by Claim~\ref{cl:ab}.
We are finally ready to define the whole path $\rho$.
Define $\rho_\up$ as $(\sigma_\up)^a$ fastened at configuration $\zeta$.
Symmetrically we define $\rho_\down$ as $(\sigma_\down)^b$ fastened
at $(\bar{q},b\cdot B+\cvalue(\bar{\zeta}))$, so that its target is $\bar{\zeta}$.
Finally, let $\rho_\rcap$ be $\sigma_\rcap$ fastened at $\target(\rho_\up)$,
and define 
$$\rho=\rho_\pref\rho_\up\rho_\rcap\rho_\down\rho_\suff.$$
Note that in this definition we did not verify properly that appropriate
sequences of transitions are fireable at certain configurations. We perform
this check in the next claim.

\begin{clm}
$\rho$ is a normal arc from $\alpha$ to $\beta$.
\end{clm}

\begin{proof}
First, condition~\ref{cl:prefsuff}\ref{ps:non-zero} ensures
that on $\rho_\pref$ all the configurations have positive counter
values apart from the source configuration $\alpha$. Similarly, on $\rho_\suff$
all the configurations have positive counter values apart from the target
configuration $\beta$. Condition~\ref{cl:prefsuff}\ref{ps:counter} asserts
that $\cvalue(\zeta)>n$, so cycle $\sigma_\up$ is fireable at $\zeta$ because
$\len(\sigma_\up)\leq n$. Since $\sigma_\up$ is a positive cycle, it can be
easily seen that also $(\sigma_\up)^a$ is fireable at $\zeta$, and moreover
on $\rho_\up$ we do not obtain any configuration with zero counter value.
A symmetric reasoning shows that $(\sigma_\down)^b$ is fireable
at $(\bar{q},b\cdot B+\cvalue(\bar{\zeta}))$ so that its target is $\bar{\zeta}$
and $\rho_\down\rho_\suff$ is a valid path with $\beta$ being the only
configuration with zero counter value.

Now observe that $\cvalue(\target(\rho_\up))=\eff(\rho_\pref)+a\cdot A$, which
is strictly larger than $L$ by condition~\ref{cl:ab}\ref{ab:inq}. Since
$\len(\sigma_\rcap)=L$ and $\source(\sigma_\rcap)=q=\state(\target(\rho_\up))$,
we see that indeed $\sigma_\rcap$ is fireable at $\target(\rho_\up)$, and
moreover on $\rho_\rcap$ all the configurations have positive counter values. 

To conclude that $\rho$ is an arc from $\alpha$ to $\beta$, it remains to
verify that $\target(\rho_\rcap)=\source(\rho_\down)$. Both these
configurations have state $\bar{q}$, so we need to verify that their counter
values are equal. However, using condition~\ref{cl:ab}\ref{ab:sum} we have
the following:
\begin{align*}
\cvalue(\target(\rho_\rcap)) & = \eff(\rho_\pref)+\eff(\rho_\up)+\eff(\rho_\rcap)\\
& = a\cdot A + K - \eff(\rho_\suff) = b\cdot B -\eff(\rho_\suff)\\
& = -\eff(\rho_\down)-\eff(\rho_\suff)=\cvalue(\source(\rho_\down)).
\end{align*}
Hence indeed $\rho$ is an arc from $\alpha$ to $\beta$,
normal by construction
(as
 $\state(\source(\rho_\rcap))$ and $\state(\target(\rho_\rcap))$
 are the base states of $\sigma_\up$ and $\sigma_\down$ respectively).\end{proof}

We summarize the properties of $\rho$ that are required in the lemma statement.
Properties~\ref{pr:up} and~\ref{pr:down} follow directly from the
construction. Properties~\ref{pr:upl} and~\ref{pr:downl} follow from our
choice of $a$ and $b$, in particular from condition~\ref{cl:ab}\ref{ab:inq}.
Property~\ref{pr:cap} follows from
condition~\ref{cl:good-cap}\ref{cap:gcd} and the fact that
$\sigma_\rcap=\ctproj(\rho_\rcap)$. Property~\ref{pr:prsu-count} follows from
condition~\ref{cl:prefsuff}\ref{ps:counter} and the fact that $\sigma_\up$
and $\sigma_\down$ are a positive and a negative cycle, respectively. Finally,
property~\ref{pr:prsu} follows from
condition~\ref{cl:prefsuff}\ref{ps:prop}. This concludes the proof of
Lemma~\ref{lem:normalization}.
\end{proof}

\subsection{Length of shortest paths}
\label{s:proof:global}

Let $\alpha$ and $\beta$ be such as in the statement of Theorem~\ref{thm:main-thm}.
Let $\rho_\circ$ be a path from $\alpha$ to $\beta$ that has the minimum possible
number of intermediate configurations
with counter value zero. Let all these intermediate configurations with counter
value zero be $\gamma_2,\ldots, \gamma_{k}$, where $\gamma_1=\alpha$ and $\gamma_{k+1}=\beta$.
For $i=1,2,\ldots,k$, let $\rho^i_\circ$ be the subpath of $\rho_\circ$ between
configurations $\gamma_i$ and $\gamma_{i+1}$. Then $\rho^i_\circ$ is an arc
from $\gamma_i$ to $\gamma_{i+1}$. By Lemma~\ref{lem:normalization}, there
exists also an arc $\rho^i$ from $\gamma_i$ to $\gamma_{i+1}$ that is either
low or is normal and admits a normal decomposition satisfying
properties \ref{pr:up}--\ref{pr:prsu}. If $\rho^i$ is low,
choose $\rho^i$ to be the shortest possible low arc from $\gamma_i$ to $\gamma_{i+1}$.
If~$\rho^i$ is normal, let
\[
\rho^i = \rho^i_\pref \, \rho^i_\up \, \rho^i_\rcap \, \rho^i_\down \, \rho^i_\suff
\]
be its normal decomposition. Our goal for the rest of the proof
(i.e., for this Section) is to show
that $\rho = \rho^1 \ldots \rho^k$, which is clearly a path from $\alpha$
to $\beta$, has length at most $14 n^2$. Note that $\rho$ has the same
number of configurations with counter value zero as~$\rho_\circ$.
Let $\Reg\subseteq \{1,2,\ldots,k\}$ be the set of indices $i$
for which $\rho^i$ is normal, and let $\Low=\{1,2,\ldots,k\}\setminus \Reg$ be
the set of indices $i$ for which $\rho^i$ is low.

First we show that the sum of the lengths of low parts of $\rho$
(more precisely, of low arcs, of $\rho^i_\pref$ and $\rho^i_\suff$)
is small. For this, Proposition~\ref{prop:basic-pumping} will be very useful.

\begin{lem}\label{lem:bound-low}
The following inequality holds:
$$
\sum_{i\in \Low} \len(\rho^i)+
\sum_{i\in \Reg} (\len(\rho^i_\pref)+\len(\rho^i_\suff))
\leq 5n^2.
$$
\end{lem}

\begin{proof}
For every $i\in \Low$, no configuration appears on $\rho^i$ more
than once, because in such a case $\rho^i$ could be made shorter using
Proposition~\ref{prop:basic-pumping} without spoiling the property that it is
low, which would contradict the assumption that $\rho^i$ is the shortest
possible. For every $i\in \Reg$, property~\ref{pr:prsu} of
Lemma~\ref{lem:normalization} ensures that no configuration appears more than
once on the paths $\rho^i_\pref$ and $\rho^i_\suff$. Suppose that some
configuration $\gamma$ appears both in $\rho^i$ and in $\rho^j$, for some $i<j$.
Then by applying Proposition~\ref{prop:basic-pumping} to configuration $\gamma$
in the path $\rho$, we would obtain a path from $\alpha$ to $\beta$ with
a strictly smaller number of intermediate configurations with counter value
zero, which would contradict our choice of $\rho_\circ$.

Hence, we conclude that among configurations appearing on paths from the set
$\{\rho^i\,\colon\, i\in \Low\}\cup \{\rho^i_\pref,\, \rho^i_\suff\,\colon\, i\in \Reg\}$,
no configuration appears more than once. Since all these paths are
low, all these configurations have counter values between $0$ and $5n-1$.
Hence, the total number of configurations appearing on these paths
is at most $n\cdot 5n=5n^2$, which concludes the proof.\end{proof}

Now we will estimate the length of the rest of the path~$\rho$. First, however, we have
to prepare a toolbox of lemmas. We introduce the following notation.
For $S,T\in \SCCs$, let $\Reg_{(S,T)}\subseteq \Reg$ be the set of all those
indices $i$ for which $\rho^i$ is $(S,T)$-normal. Moreover, let
$\Reg_{(S,\cdot)}=\bigcup_{T'\in \SCCs} \Reg_{(S,T')}$ and
$\Reg_{(\cdot,T)}=\bigcup_{S'\in \SCCs} \Reg_{(S',T)}$.

\begin{lem}\label{lem:non-repeating-caps}
Let $S,T\in \SCCs$. Suppose $i\in \Reg_{(S,\cdot)}$ and $j\in \Reg_{(\cdot,T)}$
for some $i,j$ with $1\leq i<j\leq k$.
Then there are no two configurations $\delta_i$ and $\delta_j$ appearing
on $\rho^i_\rcap$ and $\rho^j_\rcap$ respectively
such that $\state(\delta_i) = \state(\delta_j)$ and $\cvalue(\delta_i) - \cvalue(\delta_j)$
is divisible by $\gcd(\eff(\sigma^{+}_S), -\eff(\sigma^{-}_T))$.
\end{lem}

\begin{proof}
Denote $A = \eff(\sigma^{+}_S)$ and $B = -\eff(\sigma^{-}_T)$.
Assume towards contradiction that there exists $\delta_i$ on $\rho^i_\rcap$
and $\delta_j$ on $\rho^j_\rcap$ such that
$\state(\delta_i) = \state(\delta_j)$ and $\cvalue(\delta_i) - \cvalue(\delta_j)$
is divisible by $\gcd(A, B)$. 
We will show that we can modify $\rho^i$
and $\rho^j$ so that the part between $\delta_i$ and $\delta_j$
can be cut off from $\rho$.
We will then obtain a new, modified path by removing this part from it.
This will contradict the assumption that $\rho$ has the minimum
possible number of intermediate configurations
with counter value zero (see the first paragraph of Section~\ref{s:proof:global}, p.~\pageref{s:proof:global}):
indeed, the inequality $i<j$ holds by the assumptions of the lemma,
and therefore $\rho$ has at least one configuration with counter value
zero between the paths $\rho^i$ and $\rho^j$ (namely $\gamma_{i+1}$, as $i+1 \le j$). The modified path will have this configuration removed, and will
not introduce any other such configurations; thus, the number of intermediate
configurations with counter value zero in the original path is not
minimal.

Let $\cvalue(\delta_j) - \cvalue(\delta_i) = Z$,
where $Z = z \cdot \gcd(A, B)$ for some integer $z$.
Due to B\'{e}zout's identity we know that there exist $a, b \in \N$
such that $a \cdot A - b \cdot B = \gcd(A, B)$; cf. the proof of Claim~\ref{cl:ab}.
If $z \geq 0$ then $az \cdot A - bz \cdot B = z \cdot \gcd(A, B) = Z$,
where $az \geq 0$ and $bz \geq 0$.
If $z < 0$ then $(MB + az) \cdot A - (MA + bz) \cdot B = z \cdot \gcd(A, B) = Z$,
where $M$ is large enough so that $MB + az \geq 0$ and $MA + bz \geq 0$.
Therefore, there always exist numbers $a, b \geq 0$
such that $a \cdot A - b \cdot B = Z$. 

We modify the path $\rho$ as follows.
In the path $\rho^i_\up\rho^i_\rcap$ we insert $a$ cycles $\sigma^{+}_S$ at the
end of $\rho^i_\up$, and in $\rho^j_\rcap\rho^j_\down$ we insert $b$ cycles $\sigma^{-}_T$
at the front of $\rho^j_\down$.
By property~\ref{pr:prsu-count}, this insertion
does not introduce configurations with nonpositive counter values,
since each of the cycles $\sigma^{+}_S$ and $\sigma^{-}_T$ contains at most $n$~edges.
After this operation, the configuration $\delta_i$ that was originally
on $\rho^i_\rcap$ becomes lifted
to the configuration $(\state(\delta_i),\cvalue(\delta_i)+a\cdot A)$. On the other hand,
the configuration $\delta_j$ that was originally on $\rho^j_\rcap$ becomes lifted
to the configuration $(\state(\delta_j),\cvalue(\delta_j)+b\cdot B)$. However, 
$$\cvalue(\delta_i)+a\cdot A=\cvalue(\delta_j)-Z+a\cdot A=\cvalue(\delta_j)+b\cdot B.$$ 
Since $\state(\delta_i)=\state(\delta_j)$, we conclude that these two lifted
configurations are equal.
Therefore, we can perform the following operation on $\rho$:
insert the cycles $\sigma^{+}_S$ and $\sigma^{-}_T$ as described above, and cut
out the entire part of $\rho$ between the lifted configurations
originating in $\delta_i$ and $\delta_j$ by Proposition~\ref{prop:basic-pumping}.
In this manner we obtain a path from $\alpha$ to $\beta$ that has strictly less
intermediate configurations with counter value equal zero than $\rho$, which is
a contradiction.

Notice that, because of the insertions that we performed, the length
of the modified path may exceed the length of the original one;
it is only the number of intermediate configurations with counter value zero
that is guaranteed to decrease.
\end{proof}

\begin{lem}\label{lem:not-so-many-towers}
Let $S,T\in \SCCs$. Then $|\Reg_{(S,T)}|\leq \gcd(\eff(\sigma^{+}_S), -\eff(\sigma^{-}_T))$.
\end{lem}

\begin{proof}
Let us denote $A = \eff(\sigma^{+}_S)$ and $B = -\eff(\sigma^{-}_T)$.
Assume towards a contradiction that $|\Reg_{(S,T)}|>\gcd(A,B)$.
For $i\in \Reg_{(S,T)}$, let $\delta_i=\target(\rho^i_\rcap)$.
By the pigeonhole principle, for some two indices $i < j$
configurations $\delta_i$ and $\delta_j$
have the same counter value modulo $\gcd(A, B)$.
Moreover, $\delta_i$ and $\delta_j$ have the same state, which is the base state of $\sigma^{-}_T$
by our definition of a normal arc.
This contradicts Lemma~\ref{lem:non-repeating-caps}.
\end{proof}

\medskip
\subsubsection*{Total length of caps.}
We have now all the necessary ingredients to establish the desired upper bounds
on the lengths of caps.
Recall that for a strongly connected component $S\in \SCCs$
we denote by $n_S$ the number of vertices in $S$.

\begin{lem}\label{lem:sumsbounds}
Let $S,T\in \SCCs$, let $A_S=\eff(\sigma^+_S)$ and let $B_T=-\eff(\sigma^-_T)$. Then:
\begin{align}
\sum_{i\in \Reg_{(S,\cdot)}} \len(\rho^i_\rcap)&\leq A_S\cdot n;\label{bnd:S.}\\
\sum_{i\in \Reg_{(\cdot,T)}} \len(\rho^i_\rcap)&\leq n\cdot B_T;\label{bnd:.T}\\
\text{\llap{moreover,\quad}}
\sum_{i\in \Reg} \len(\rho^i_\rcap)&\leq n^2.\label{bnd:..}
\end{align}
\end{lem}

\begin{proof}
For~\eqref{bnd:S.}, assume towards a contradiction
that $\sum_{i\in \Reg_{(S,\cdot)}} \len(\rho^i_\rcap)>A_S\cdot n$.
Then by the pigeonhole principle there exists
two configurations $\delta$ and $\delta'$ on the paths $\rho^i_\rcap$
for $i\in \Reg_{(S,\cdot)}$ which have the same state and the same counter
value modulo $A_S$.
Assume w.l.o.g. that $\delta$ is earlier in the path than $\delta'$.
By property~\ref{pr:cap} of Lemma~\ref{lem:normalization},
configurations $\delta$ and $\delta'$ cannot
appear in the same path $\rho^i_\rcap$. Indeed, otherwise the projection of the
part of $\rho^i_\rcap$ between $\delta$
to $\delta'$ would be a cycle with effect divisible by $A_S$,
so also by $\gcd(A_S, -\eff(\sigma^{-}_T))$,
where $T$ is the SCC for which $\rho^i$ is $(S,T)$-normal.
Therefore they have to belong to different arcs.
Let $\delta$ belong to $\rho^i$ and $\delta'$ belong to $\rho^j$,
where $j\in \Reg_{(S,T)}$ for some $T\in \SCCs$.
However, by Lemma~\ref{lem:non-repeating-caps},
there are no two configurations $\delta$ and $\delta'$ on $\rho^i$ and $\rho^j$,
respectively, such that their states are the same and the difference
in counter values is divisible by $\gcd(A_S, -\eff(\sigma^{-}_T))$.
Contradiction, as $\delta$ and $\delta'$ are such configurations:
the difference of its counter values is divisible by $A_S$,
so also by $\gcd(A_S, -\eff(\sigma^{-}_T))$. Thus~\eqref{bnd:S.} is proved,
and~\eqref{bnd:.T} follows from a symmetric reasoning.
The bound~\eqref{bnd:..} follows by summing~\eqref{bnd:S.}
through all $S\in \SCCs$ and using the facts that
$\eff(\sigma^{+}_S)\leq n_S$ and $\sum_{S\in \SCCs} n_S=n$.\end{proof}

\medskip
\subsubsection*{Total length of positive and negative cycles.}
We now show that the total sum of the lengths of $\rho^i_\up$
and $\rho^i_\down$ is at most $8 n^2$.
This is the case where we need
the key estimations~\ref{pr:upl} and~\ref{pr:downl} in
Lemma~\ref{lem:normalization}.

\begin{lem}\label{lem:sumcycles}
The following inequalities hold:
\begin{align*}
\sum_{i \in \Reg} \len(\rho^i_\up) \leq 4n^2& ,\qquad &\sum_{i \in \Reg} \len(\rho^i_\down) \leq 4n^2.
\end{align*}
\end{lem}

\begin{proof}
We show how to bound the sum of lengths of paths $\rho^i_\up$.
For any $S\in \SCCs$, let us denote $A_S = \eff(\sigma^{+}_S)$
and $B_S = -\eff(\sigma^{-}_S)$.
For each $i\in \Reg$, let $S_i,T_i\in \SCCs$ be such that
$\rho^i$ is $(S_i, T_i)$-normal, and let $L_i = \len(\rho^i_\rcap)$.
By Lemma~\ref{lem:normalization} we know that
$\eff(\rho^i_\up) \leq 2L_i + 2\cdot \lcm(A_{S_i}, B_{T_i})$.
Since $\ctproj(\rho^i_\up)=(\sigma^{+}_{S_i})^a$ for some integer $a$,
we have
$$
\len(\rho^i_\up)
=
\eff(\rho^i_\up) \cdot \frac{\len(\sigma^{+}_{S_i})}{\eff(\sigma^{+}_{S_i})}
\leq
\eff(\rho^i_\up) \cdot \frac{n_{S_i}}{A_{S_i}}
\leq
(2L_i + 2\cdot \lcm(A_{S_i}, B_{T_i})) \cdot \frac{n_{S_i}}{A_{S_i}}.
$$
Hence,
\begin{equation}\label{eq:total-length}
\sum_{i \in \Reg} \len(\rho^i_\up) \leq 2 \sum_{i \in \Reg} \frac{L_i\, n_{S_i}}{A_{S_i}}
+ 2 \sum_{i \in \Reg} \frac{\lcm(A_{S_i}, B_{T_i}) \cdot n_{S_i}}{A_{S_i}}.
\end{equation}
We will separately estimate the first and the second term.
First we focus on $\sum_{i \in \Reg} \frac{L_i\, n_{S_i}}{A_{S_i}}$.
Let us fix some specific $S\in \SCCs$.
We have
\[
\sum_{i \in \Reg_{(S,\cdot)}} \frac{L_i\, n_{S_i}}{A_{S_i}}
=
\frac{n_S}{A_S}\cdot \sum_{i \in \Reg_{(S,\cdot)}} L_i
\leq
\frac{n_S}{A_S} \cdot A_S \cdot n = n_S \cdot n,
\]
where the inequality follows from Lemma~\ref{lem:sumsbounds}\eqref{bnd:S.}.
Thus 
\begin{equation}\label{eq:first-factor}
\sum_{i \in \Reg} \frac{L_i\, n_{S_i}}{A_{S_i}}
=
\sum_{S \in \SCCs}\, \sum_{i \in \Reg_{(S,\cdot)}} \frac{L_i\, n_{S_i}}{A_{S_i}}
\leq
\sum_{S \in \SCCs} n_S \cdot n = n^2.
\end{equation}
In order to estimate the second term, fix some $S,T\in \SCCs$.
Note that
$\frac{\lcm(x, y)}{x} = \frac{xy}{\gcd(x,y) \cdot x} = \frac{y}{\gcd(x, y)}$
for all positive integers $x,y$.
Now we have
\begin{align*}
\sum_{i \in \Reg_{(S,T)}} \frac{B_{T_i} \cdot n_{S_i}}{\gcd(A_{S_i}, B_{T_i})}
=
\sum_{i \in \Reg_{(S,T)}} \frac{B_T \cdot n_S}{\gcd(A_S, B_T)}
=
|\Reg_{(S,T)}| \cdot \frac{B_T \cdot n_S}{\gcd(A_S, B_T)} \\
\leq
\gcd(A_S, B_T) \cdot \frac{B_T \cdot n_S}{\gcd(A_S, B_T)}
=
B_T \cdot n_S \leq n_T \cdot n_S,
\end{align*}
where the first inequality follows from Lemma~\ref{lem:not-so-many-towers}
and the second one from the fact that the effect of a path is bounded by its length.
Therefore,
\begin{eqnarray}\label{eq:second-factor}
\sum_{i \in \Reg} \frac{B_{T_i} \cdot n_{S_i}}{\gcd(A_{S_i}, B_{T_i})}
=
\sum_{S, T \in \SCCs}\, \sum_{i \in \Reg_{(S, T)}}
\frac{B_{T_i} \cdot n_{S_i}}{\gcd(A_{S_i}, B_{T_i})} \nonumber \\
\leq
\sum_{S, T \in \SCCs} n_T \cdot n_S
=
\sum_{S \in \SCCs} n_S \cdot \sum_{T \in \SCCs} n_T
=
n^2.
\end{eqnarray}
By connecting equations \eqref{eq:total-length}, \eqref{eq:first-factor}
and \eqref{eq:second-factor} we obtain
\[
\sum_{i \in \Reg} \len(\rho^i_\up) \leq 2n^2 + 2n^2 = 4n^2.
\]
The upper bound on the sum of lengths of paths $\rho^i_\down$
is obtained
analogously, using Lemma~\ref{lem:sumsbounds}\eqref{bnd:.T}
instead of Lemma~\ref{lem:sumsbounds}\eqref{bnd:S.}.\end{proof}

\noindent
Combining the bounds of Lemma~\ref{lem:bound-low},
Lemma~\ref{lem:sumsbounds}\eqref{bnd:..}, and Lemma~\ref{lem:sumcycles},
we conclude that
\[
\len(\rho) \leq 5 n^2 + n^2 + 4 n^2 + 4 n^2 \leq 14 n^2,
\]
which completes the proof of Theorem~\ref{thm:main-thm}.


\section{Generalizations}
\label{s:generalizations}

\subsection{Proof of Theorem~\ref{thm:generalization}}

In this section we prove Theorem~\ref{thm:generalization}, which provides an upper bound on the length of the shortest path between any pair of configurations. 
For convenience, we recall its statement.
\restategeneralization*

\begin{proof}
Let $a = \cvalue(\alpha)$ and $b = \cvalue(\beta)$.
Assume without loss of generality that $a \geq b$; the second case is symmetric.
Let $\rho$ be some path from $\alpha$ to $\beta$. 
We first formulate the following claim.

\begin{clm}\label{cl:lifted-automaton}
Let $a \ge 0$ be chosen arbitrarily; then
there exists an OCS $\mathcal{O}^a$ with the following property.
For any $p, q \in Q$ the OCS $\mathcal{O}^a$ has a path from $(p, 0)$ to $(q, 0)$ of length exactly $K$
(i.e., with $K+1$ configurations)
if and only if the OCS $\mathcal{O}$ has a path from $(p, a)$ to $(q, a)$
which contains exactly $K+1$ configurations
of counter value at least~$a$
(and possibly other configurations).
Moreover, $\mathcal{O}^a$ and $\mathcal{O}$ have the same number of states.
\end{clm}

\begin{proof}
We construct $\mathcal{O}^a$ with the set of states $Q^a$, the set of non-zero transitions $T_{>0}^a$
and the set of zero tests $T_{=0}^a$ as follows. It has the same set of states as $\mathcal{O}$,
so $Q^a = Q$, and the same set of non-zero transitions, so $T_{>0}^a = T_{>0}$.
Only the set of zero tests is different.
The set $T_{=0}^a$ contains only tuples of the form $(q, 0, q')$ for $q, q' \in Q^a$. A tuple $(q, 0, q')$ belongs to $T_{=0}^a$
if and only if there is a path in $\mathcal{O}$ from configuration $(q, a)$ to configuration $(q', a)$ such that
all the intermediate configurations have counter value smaller than $a$.

We now verify that this OCS $\mathcal{O}^a$ satisfies our requirements.

Suppose there is a path from $(p, 0)$ to $(q, 0)$ in $\mathcal{O}^a$.
Observe that there is a corresponding path from $(p, a)$ to $(q, a)$ in $\mathcal{O}$.
Every non-zero transition from $(r, c)$ for $c > 0$ in $\mathcal{O}^a$ is simulated by one transition from $(r, c+a)$ in $\mathcal{O}$, as $T_{>0}^a = T_{>0}$.
Every zero-test from $(r, 0)$ to $(r', 0)$ in $\mathcal{O}^a$ is simulated by a path from $(r, a)$ to $(r', a)$ in $\mathcal{O}$, where all
the intermediate configurations have counter value smaller than $a$. Such a path exists by the definition of $T_{=0}^a$.
Note that the corresponding path of $\mathcal{O}$ indeed has exactly as many configurations with counter value at least $a$
as there are configurations in the original path of $\mathcal{O}^a$.

Now suppose there is a path from $(p, a)$ to $(q, a)$ in $\mathcal{O}$.
We construct the corresponding path of $\mathcal{O}^a$ as follows. Every part of the path from some $(r, a)$
to some $(r', a)$ where all the configurations in between have smaller counter values is replaced by a zero-test 
of $\mathcal{O}^a$. The constructed path of $\mathcal{O}^a$ indeed has as many configurations
as there are configurations in the path of $\mathcal{O}$ which have counter value at least $a$.\end{proof}

Let $\gamma = (q, a)$ be the last configuration in $\rho$
which has counter value $a$. Suppose $\alpha = (p, a)$ and
consider the OCS $\mathcal{O}^a$ from Claim~\ref{cl:lifted-automaton}.
As there is a path from $(p, a)$ to $(q, a)$ in $\mathcal{O}$ then there is a path
from $(p, 0)$ to $(q, 0)$ in $\mathcal{O}^a$. By Theorem~\ref{thm:main-thm} there is
a path from $(p, 0)$ to $(q, 0)$ of length at most $14 n^2$. Now one more time by
Claim~\ref{cl:lifted-automaton} there is a path from $(p, a)$ to $(q, a)$ which has
at most $14 n^2 + 1$ configurations of counter value at least $a$. Let us denote by $\rho'$
the concatenation of this path and the suffix of $\rho$ that starts in $\gamma$ and finishes in $\beta$.
Since $\gamma$ is the last configuration in $\rho$ which has counter value at least $a$
then also in $\rho'$ there are at most $14n^2 + 1$ configurations of counter value at least $a$.

Let $\rho''$ be a shortest path from $\alpha$ to $\beta$ such that there are at most
$14n^2 + 1$ configurations in this path with counter value at least $a$. There is at least one
such path, namely $\rho'$, so a shortest one clearly exists.
Let us define a set $\low$ as the set of all configurations appearing in $\rho''$ whose counter values are smaller than $a$.
We claim that $|\low| \leq an$. Indeed, assume the converse, i.e., $|\low| > an$. Then, by the pigeonhole principle,
some two configurations appearing on $\rho''$, say $\delta$ and $\delta'$, are equal. 
Then by Proposition~\ref{prop:basic-pumping}, cutting out the part of the path between $\delta$ and $\delta'$ would leave 
a strictly shorter path from $\alpha$ to $\beta$ that would have not more configurations with counter value at least $a$. 
This would contradict our choice of~$\rho''$.

We therefore know that there are at most $14n^2+1$ configurations in $\rho''$ with counter value at least $a$
and at most $an$ configurations with counter value smaller than $a$. Thus altogether there are at most $14n^2+1+an$
configurations on $\rho''$, so $\len(\rho'') \leq 14n^2 + an \leq 14n^2 + n \cdot \max(a,b)$. This completes the proof
of Theorem~\ref{thm:generalization}.\end{proof}

\subsection{Generalization to counters with values in $\mathbb{Z}$}

\newcommand{\negu}{{\textrm{neg}}}

Furthermore, in this section we show how our results can be used to give improved upper
bounds on the length of the shortest path in the model considered by Alur and
\v{C}ern\'y~\citep{AlurC11}. Recall that in this model, the counter can take
arbitrary values in~$\mathbb Z$ and there are no zero-tests. In fact, we can
show that a quadratic upper bound holds in a much more general model, where
zero tests are allowed and transitions fireable at positive counter values may
differ from transitions fireable at negative counter values. We start with
defining formally the model we are working with.

A \emph{one-$\mathbb{Z}$-counter system} (\emph{$\mathbb{Z}$-OCS}) \OCS
consists of a finite set of \emph{states} $Q$,
a set of \emph{positive transitions} $T_{>0} \subseteq Q \times \{-1,0,1\} \times Q$,
a set of \emph{negative transitions} $T_{<0} \subseteq Q \times \{-1,0,1\} \times Q$,
and a set of \emph{zero tests} $T_{=0} \subseteq Q \times \{-1,0, 1\} \times Q$.
The set of \emph{transitions} is $T = T_{>0}\cup T_{<0} \cup T_{=0}$. The positive
transitions are fireable in configurations where the counter value is positive,
negative transitions are fireable whenever the counter value is negative, and
zero tests are fireable whenever the counter value is equal to zero. We adopt
all the notation from one-counter systems in a natural way. In particular, the
configurations of a $\mathbb{Z}$-OCS \OCS are pairs $(q,c)$, where $q\in Q$ is
the configuration's state, and $c\in \mathbb{Z}$ is the configuration's counter
value. Observe that one-$\mathbb{Z}$-counter systems generalize standard
one-counter systems, because we can take $T_{<0}=\emptyset$ and disallow zero
tests having effect $-1$ on the counter.

Again, a path is a sequence of the form 
\[
(\gamma_1, t_1) (\gamma_2, t_2) \ldots (\gamma_m, t_m) \in ((Q \times \mathbb{Z}) \times T)^*,
\]
for which some final configuration $\gamma_{m+1}$ exists, such that for each
$i=1,2,\ldots,m$ we have that $\gamma_i \trans{t_i} \gamma_{i+1}$, i.e., firing
transition $t_i$ at configuration $\gamma_i$ results in configuration
$\gamma_{i+1}$. We now state formally our result for one-$\mathbb{Z}$-counter
systems.

\begin{thm}\label{thm:Z-generalization}
Let \OCS be a one-$\mathbb{Z}$-counter system with $n$ states.
Suppose configuration $\beta$ is reachable
from configuration $\alpha$ in \OCS,
where $\cvalue(\alpha) = c_\alpha$ and $\cvalue(\beta) = c_\beta$.
Then \OCS has a path from $\alpha$ to $\beta$
of length at most $56 n^2 + 2n \cdot \max(|c_\alpha|,|c_\beta|)$.
\end{thm}

We remark that one can approach Theorem~\ref{thm:Z-generalization} by following
the lines of the proofs of Theorems~\ref{thm:main-thm}
and~\ref{thm:generalization}, and adjusting the argumentation to the setting of
one-$\mathbb{Z}$-counter systems. The proof, however, would be even more
technical, because having both positive and negative counter values requires
performing the pumping arguments twice: both for very high (positive) values
and for very low (negative) values. Instead, we show how
Theorem~\ref{thm:Z-generalization} can be deduced from
Theorem~\ref{thm:main-thm}.

\begin{proof}
The summary of the argument is as follows.
We rely on a construction of a one-counter system that faithfully
simulates the given one-$\Z$-counter system.
Observe that zero test transitions let us
maintain, in the finite control state, information about whether the counter value is positive.
Then the counter itself can be used to store the absolute value only.
The obtained system will be twice as big as the original one,
which will give us the required bounds.

In more detail, 
let $\mathcal{O}$ be the given $\mathbb{Z}$-OCS, and let $Q$, $T_{>0}$, $T_{<0}$,
and $T_{=0}$ be its
set of states, its sets of positive and negative transitions, and its set of
zero tests, respectively. 
%
%
%
%
Define a (standard) one counter system $\mathcal{O}^+$ as follows:
it has states $Q^+=Q\times\{+,-\}$ and sets of non-zero and zero transitions
$T_{>0}^+$ and $T_{=0}^+$. We set
\begin{align*}
T_{>0}^+=
&\{ ((q,+), c, (q',+))\text{\ for all\ }(q,c,q')\in T_{>0}\}\cup {}\\
&\{ ((q,-), -c, (q',-))\text{\ for all\ } (q,c,q')\in T_{<0}\}
\quad\text{and}\\
T_{=0}^+=
&\{ ((q,+), 1, (q',+))\text{\ for all\ }(q,1,q')\in T_{=0}\} \cup {}\\
&\{ ((q,+), 0, (q',+))\text{\ for all\ }(q,0,q')\in T_{=0}\} \cup {}\\
&\{ ((q,-), 1, (q',-))\text{\ for all\ }(q,-1,q')\in T_{=0}\} \cup {}\\
&\{ ((q,+), 0, (q,-))\text{\ for all\ }q\in Q\} \cup {}\\
&\{ ((q,-), 0, (q,+))\text{\ for all\ }q\in Q\}.
\end{align*}
Finally, let $\alpha^+=((\state(\alpha),\sign(\cvalue(\alpha))), |\cvalue(\alpha)|)$ and 
$\beta^+=((\state(\beta),\sign(\cvalue(\beta))), |\cvalue(\beta)|)$ where $\sign(x)$
is $+$ or $-$ depending on whether the number~$x$ is positive or not.

Now $\mathcal{O}^+$ is a one-counter system with $2 n$ states.
By Theorem~\ref{thm:main-thm}, the 
length of the shortest path from $\alpha^+$ to $\beta^+$ in $\mathcal{O}^+$ is bounded by
$14\cdot (2 n)^2 + 2 n \cdot \max(\cvalue(\alpha^+),\cvalue(\beta^+))=56 n^2 + 2 n \cdot \max(|\cvalue(\alpha)|,|\cvalue(\beta)|)$.
Thus, it suffices to show that the system $\mathcal{O}^+$ has a path from
$\alpha^+$ to $\beta^+$ if and only if
the system $\mathcal{O}$ has a path from
$\alpha$ and $\beta$; and, moreover, that if these paths exist,
then the shortest such path in $\mathcal{O}$ is at most as long
as the shortest such path in $\mathcal{O}^+$.

It is immediate from the construction of $\mathcal{O}^+$ that
every path in $\mathcal{O}^+$ corresponds to a path in $\mathcal{O}$
and vice versa.
Indeed, define a function $\varphi$ that maps any path in $\mathcal{O}^+$ from $\alpha^+$ to $\beta^+$
into a path in $\mathcal{O}$ from $\alpha$ to $\beta$ as follows.
First let $\varphi$ be defined on individual states and transitions by
\begin{align*}
\varphi\left((q, +)\right)&=
\varphi\left((q, -)\right)=
q,\\
\varphi\left((q,+),0,(q,-)\right)&=
\varphi\left((q,-),0,(q,+)\right)=
\begin{cases}
\varepsilon &\text{if $(q, 0, q) \not\in T_{=0}$ \quad and}\\
(q,0,q) &\text{otherwise},
\end{cases}\\
\varphi\left((q,+),c,(q',+)\right)&=(q,c,q'), \quad\text{and}\\
\varphi\left((q,-),c,(q',-)\right)&=(q,-c,q').
\end{align*}
Then extending $\varphi$ to a homomorphism with respect to the concatenation of
paths will give us the required correspondence,
$\varphi$ will be onto,
and for every path $\rho^+$ in $\mathcal{O}^+$ we will have
$\len(\phi(\rho^+))\leq \len(\rho^+)$. This completes the proof.
\end{proof}

Notice that the model used by Alur and \v{C}ern\'y~\citep{AlurC11}
corresponds to setting $T_{>0}=T_{<0}=T_{=0}$ in our definition of
a one-$\mathbb{Z}$-counter system. In this case, one can very easily obtain
from the statement of Theorem~\ref{thm:Z-generalization}
a marginally better upper bound of $56n^2+ 2 n \cdot |c_\alpha-c_\beta|$. Indeed,
since the fireability of transitions of $\mathcal{O}$ is independent of the
sign of the counter, on any path in $\mathcal{O}$ we can add an arbitrary
integer to all the counter values throughout the path, and we still obtain a path
of $\mathcal{O}$. Hence, by decrementing all the counter values by $c_\alpha$,
we can equivalently consider the problem of bounding the length of the shortest
path from $\alpha$ to $\beta$ when we know that $c_\alpha=0$. Then an
application of Theorem~\ref{thm:Z-generalization} yields a path from $\alpha$ to
$\beta$ with length at most $56n^2+ 2 n \cdot |c_\beta|$, which translates to the
bound $56n^2+ 2 n \cdot |c_\alpha-c_\beta|$ in the general case before
decrementing.

\section{Pumping lemma for one-counter languages}
\label{s:pumping-constant}

What else can be obtained using our technique
and what are its limits?
In this section we explain how our work relates
to the classic ``pumping'' technique for formal languages.
%
%
It is not difficult to see that our proofs develop
an advanced version of this technique, which we use for ``unpumping'',
or ``pumping down'' the path that traverses the configuration graph
of the automaton. Remarkably, however,
our arguments do not immediately deliver any improved
upper bounds on the value of the so-called \df{pumping constant}.

Indeed, consider the following forms of the \df{pumping lemma},
stated for one-counter automata (OCA):
For every OCA~\OCA
there exist nonnegative constants $N_1, \ldots, N_4$
for which the following statements hold (for $i = 1, 2, 3, 4$, respectively):
each word $w$ of length at least~$N_i$ in the language $L$ of \OCA
has a decomposition $w = x u y v z$ with $\len(u) + \len(v) > 0$
such that
\begin{equation*}
\begin{array}{ll}
(i = 1) & x y z \in L, \\
(i = 2) & x u^n y v^n z \in L \text{\ for all\ } n \ge 1, \\
(i = 3) & x u^n y v^n z \in L \text{\ for all\ } n \ge 0, \\
(i = 4) & \text{for some $k, \ell \ge 1$ it holds that
$x u^{1 + k n} y v^{1 + \ell n} z \in L$ for all $n \ge 0$.}
\end{array}
\end{equation*}
Form~3 is the usual (form of a) pumping lemma, and $N_3$ is the usual
pumping constant. Sometimes the lemma is stated in the form~2,
as in Latteux~\cite{Latteux83}; in this particular case, the proof can in fact
be used to show the stronger form~3. One should not probably refer
to form~1 as a pumping lemma; it is rather an ``unpumping'' or ``downpumping''
lemma, the ``complement'' of form~2 with respect to form~3.
Form~4, in comparison, does pump the word ``up'', but it
may require iterating infixes $u$ and $v$ several times (the number
of iterations can be different for $u$ and $v$, $k \ne \ell$)
in order for the word to satisfy form~2
(cf.~our proof of Lemma~\ref{lem:normalization}).

Suppose,
for an individual OCA~\OCA with $n$~states,
that $N_i$ are
the \emph{smallest numbers} satisfying the requirements above.
Then the following statements hold.
\begin{clm}
\label{c:disc1}
$N_3 = \max(N_1, N_2)$.
\end{clm}
\begin{clm}
\label{c:disc2}
$N_2 \ge N_4$.
\end{clm}
\begin{clm}
\label{c:disc3}
There is an OCA that has
$N_i = \MyOmega(n^2)$ for all $i$.
\end{clm}
Claims~\ref{c:disc1} and~\ref{c:disc2} are immediate,
and Claim~\ref{c:disc3} is justified by examples in Section~\ref{s:overview}.
Previous techniques
show that all $N_i$ are at most cubic in~$n$, 
so there is a familiar gap between $n^2$ and $n^3$.
\begin{clm}
\label{c:disc4}
$N_4 = O(n^2)$.
\end{clm}
\begin{proof}
If a word $w \in L$ has length
quadratic or larger, then either some configuration along the accepting path
repeats, with the result that we can choose $v = \epsilon$ and $k = 1$,
or the maximum counter value along the path is at least $n + 1$,
and so there is a pair of cycles with effects $A > 0$ and $- B < 0$
that we can repeat $k = B$ and $\ell = A$ times, respectively.
If the path in $\OCA$ accepts, but finishes in a configuration
with a nonzero counter value,
then there might be no need to find any negative cycle at all.
\end{proof}
\begin{rem}
This argument resolves the associated \df{longest accepted word}
problem: if a language $L$ of an OCA \OCA is finite, then no word in~$L$
has super-quadratic length.
\end{rem}
The optimal choice of $N_1$, $N_2$, and $N_3$ presents an open
problem. For $N_1$, the reason that the technique from the present paper fails to deliver
a quadratic upper bound is as follows: we not only remove fragments
of the path, but also, crucially, insert several additional copies of positive
and negative cycles, as well as auxiliary paths in the middle of the computation.
In fact, our core ``unpumping modulo $\gcd(\cdot, \cdot)$'' technique may need
to first insert fresh copies of cycles in order to shorten the path.
This goes beyond subsequence-oriented unpumping and seems to be incompatible
with it. In fact, it is not known if an even weaker version of the pumping
lemma, one that permits arbitrary removal of subpaths, allows a subcubic
pumping constant. A positive answer to this question could lead to a new
proof of our main result.

\section{Open problems}
\label{s:conc}

In conclusion, we have shown that
any one-counter automaton with $n$ states, unless its language is
empty, accepts some word of length at most $14 n^2$.
This closes the gap between the previously known upper bound of $O(n^3)$ and
lower bound of $\MyOmega(n^2)$, strengthening results that have previously
appeared in the literature.
Our treatment of automata with zero tests uses a ``global'' argument
on paths (computations of one-counter automata).
Our techniques also provide a tight upper bound on the length of shortest paths
between arbitrary configurations in one-counter transition systems,
both in the model where the counter stays nonnegative,
and in the model where it can take arbitrary values from $\mathbb Z$.

We note one open problem of particular interest:
\begin{enumerate}
\item
To demonstrate that language equivalence for \emph{deterministic}
one-counter automata can be decided in nondeterministic logspace ($\mathbf{NL}$),
B\"ohm et~al.~\cite{BohmGJ13} prove the following result:
if the languages of any two one-counter automata~$\OCA_1$ and $\OCA_2$ are
different, then there is a word $w$ in their symmetric difference
that has length at most $p(n)$, polynomial in~$n$ (the maximum of
the number of states of $\OCA_1$ and $\OCA_2$).
Finding tight bounds on $p(n)$ is an open problem.
Our results show that $p(n)$ need not be superquadratic
if the language of $\OCA_2$ is empty; we do not even require $\OCA_1$
to be deterministic. Note that language equivalence is undecidable
if at least one of $\OCA_1$ and $\OCA_2$ is nondeterministic
(because language universality is undecidable and a special case of this problem),
so the length of the shortest distinguishing word for a nondeterministic
OCA and a deterministic finite automaton cannot have any a priori upper bound.
\end{enumerate}
The shortest path question can be considered for
transition systems of other models of computation. For pushdown
automata with $n$ states, a binary stack alphabet, and transitions that
push/pop individual stack symbols only,
it is not difficult to see that
shortest paths have length $2^{O(n^2)}$,
with a worst-case lower bound of $2^{\MyOmega(n)}$.
(A related question for valence automata over the free group
was studied by Ang et~al.~\cite{AngPRS09}.)
In fact, the tight bounds are $2^{\Theta(n^2 / \log n)}$,
see Pierre~\cite{P92}.
In addition to the problem above,
there are also questions related to shortest paths for
one-counter automata and one-counter systems that we leave open:
\begin{enumerate}
\setcounter{enumi}{1}
\item
One can further shrink the gap between upper and lower bounds,
obtaining better estimates of the constant factor.
In the setting with zero tests, the gap is between
$n^2 / 2 - O(n)$ and $14 n^2$.
\item
If additional properties of the system are
known, stronger upper bounds may be obtained.
What properties entail subquadratic or linear
shortest paths?
\item
Our results prove the existence of paths of length $O(n^2)$;
how efficiently can such paths be found? Bradford~\cite{Bradford18}
gives a sketch of an approach more efficient
than a search in the cubic-size graph of all configurations with
counter value at most $n^2$, achieving running time $n^\omega \cdot \polylog(n)$,
where $\omega < 2.373$ is the matrix multiplication exponent.
\item
Is there a quadratic upper bound for the pumping constant
for one-counter languages? See Section~\ref{s:pumping-constant}
for discussion.
\end{enumerate}


\section*{Acknowledgment}
\noindent We are grateful to Christoph Haase, Aditya Kanade, and Georg Zetzsche
for discussions and comments.
We would also like to thank Alexander Rubtsov and Mikhail Vyalyi for bringing the papers by Del\'eage and Pierre~\cite{DP86} and Pierre~\cite{P92} to our attention.

\bibliographystyle{abbrvnat}
\bibliography{citat}

\end{document}